\documentclass[%
reprint,
superscriptaddress,
jor,
 amsmath,amssymb,
 aps,
floatfix,
]{revtex4-1}
\usepackage{amsmath}
\usepackage{marvosym}
\usepackage{graphicx,subfigure}         
\usepackage{amsmath} 
\usepackage{amssymb,stmaryrd}   
\usepackage{setspace}
\usepackage{color}
\usepackage{gensymb}
\usepackage{wrapfig}
\usepackage{mathtools}

\usepackage{flushend}
\usepackage{tikz}

\usepackage{natbib}
\newcommand{\upcite}[1]{\setcitestyle{super}\cite{#1}}

\usepackage{comment} 
\usepackage{siunitx}

\usepackage{appendix}
\usepackage{hyperref}
\usepackage{textcomp}

\DeclareUnicodeCharacter{2061}{}

\usepackage{gensymb}
\usepackage[utf8]{inputenc}

\usepackage{graphicx} 
\usepackage{marginnote}

\hypersetup{colorlinks=true, linkcolor=blue, citecolor=blue, urlcolor=blue,}

\usepackage{ragged2e}
\usepackage{setspace}
\usepackage{lettrine}

\begin{document}

\makeatletter
\renewcommand{\maketitle}{\bgroup\setlength{\parindent}{10pt}
\begin{flushleft}
  \textbf{\@title}

  \@author
\end{flushleft}\egroup
}
\makeatother

\onecolumngrid
\title{\textsf{\huge {Non-volatile spin transport in a single domain multiferroic}}}

\vspace{20cm}
\date{}
\author{%
\textsf{\textbf{Sajid Husain$^{1,	\dagger,*}$, Isaac Harris$^{1,2,\dagger,}$, Peter Meisenheimer$^{3}$, Sukriti Mantri$^{4}$, Xinyan Li$^{5}$ Maya Ramesh$^{6}$, Piush Behera$^{1,3}$, Hossein Taghinejad$^{2,7}$, Jaegyu Kim$^{3}$, Pravin Kavle$^{1,3}$, Shiyu Zhou$^{8}$, Tae Yeon Kim$^{3}$, Hongrui Zhang$^{1,3}$, Paul Stephenson$^{9}$, James G. Analytis$^{2}$, Darrell Schlom$^{6}$, Sayeef Salahuddin$^{3,10}$, Jorge Íñiguez-González$^{11,12}$, Bin Xu$^{13}$, Lane W. Martin$^{1,3,14,15}$, Lucas Caretta$^{8,16}$, Yimo Han$^{5}$, Laurent Bellaiche$^{4}$, Zhi Yao $^{1,*}$, Ramamoorthy Ramesh $^{1,2,3,14,*}$\\}
\textit{$^{1}$Materials Science Division, Lawrence Berkeley National Laboratory, Berkeley, CA, 94720, USA\\
$^{2}$Department of Physics, University of California, Berkeley, CA, 94720, USA\\
$^{3}$Department of Materials Science and Engineering, University of California, Berkeley, CA, 94720, USA\\
$^{4}$Physics Department and Institute for Nanoscience and Engineering, University of Arkansas, Fayetteville, Arkansas 72701, USA\\
$^{5}$Materials Science and NanoEngineering, Rice University, Houston, Texas, 77005, USA.\\
$^{6}$Department of Materials Science and Engineering, Cornell University, Ithaca, NY, 14850, USA\\
$^{7}$Heising-Simons Junior Fellow, Kavli Energy NanoScience Institute (ENSI), University of California, Berkeley, CA, 94720, USA\\
$^{8}$Department of Physics, Brown University, Providence, RI, 02906, USA\\
$^{9}$Department of Physics, Northeastern University, Boston, MA, 02115, USA\\
$^{10}$Department of Electrical Engineering and Computer Sciences, University of California, Berkeley, CA 94720, USA\\
$^{11}$Department of Materials Research and Technology, Luxembourg Institute of Science and Technology, 5 Avenue des Hauts-Fourneaux, L-4362 Esch/Alzette, Luxembourg\\
$^{12}$Department of Physics and Materials Science, University of Luxembourg, 41 Rue du Brill, L-4422 Belvaux, Luxembourg\\
$^{13}$Institute of Theoretical and Applied Physics, Jiangsu Key Laboratory of Thin Films, School of Physical Science and Technology, Soochow University, Suzhou 215006, China\\
$^{14}$Departments of Materials Science and 
NanoEngineering, Chemistry, and Physics 
and Astronomy, Rice University, Houston, TX, 77005, 
USA.\\
$^{15}$Rice Advanced Materials Institute, Rice 
University, Houston, TX, 77005, USA.\\
$^{16}$School of Engineering, Brown University, Providence, RI, 77005, USA\\}
{$^{*}$rramesh@berkeley.edu}\\
{$^{*}$jackie-zhiyao@lbl.gov}\\
{$^{*}$shusain@lbl.gov}\\
{$^\dagger$ These authors contributed equally}
}}

\date{\today}
\maketitle

\textbf{\large{Antiferromagnets have attracted significant attention in the field of magnonics, as promising candidates for ultralow-energy carriers for information transfer for future computing. The role of crystalline orientation distribution on magnon transport has received very little attention.  In multiferroics such as BiFeO$_3$ the coupling between antiferromagnetic and polar order imposes yet another boundary condition on spin transport. Thus, understanding the fundamentals of spin transport in such systems requires a single domain, a single crystal. We show that through Lanthanum(La) substitution, a single ferroelectric domain can be engineered with a stable, single-variant spin cycloid, controllable by an electric field.  The spin transport in such a single domain displays a strong anisotropy, arising from the underlying spin cycloid lattice. Our work shows a pathway to understand the fundamental origins of spin transport in such a single domain multiferroic.}}
\\
\twocolumngrid
\setcounter{figure}{0}
\renewcommand{\figurename}{\textbf{Figure}}

\begin{figure*}[htbp!]
\centering
\includegraphics[width=18cm]{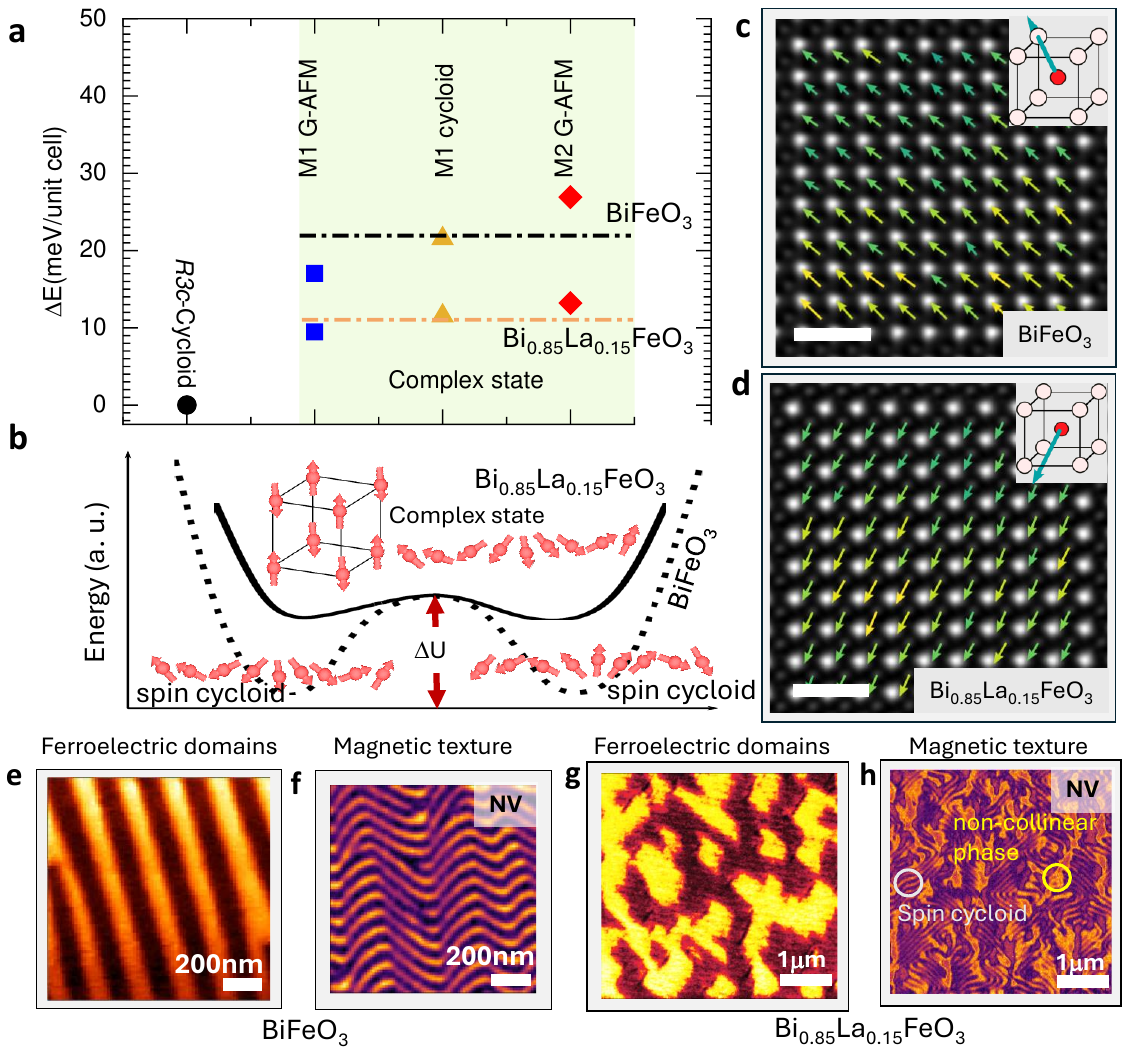}
\caption{\label{fig:structure} \textbf{Ferroelectric and magnetic ground state of La substituted BiFeO$_3$:} Effective Hamiltonian calculated \textbf{a} magnetic ground state energy of the spin cycloid and G-type antiferromagnetic phase in BiFeO$_3$ and Bi$_{0.85}$La$_{0.15}$FeO$_3$. $R3c$ represents uniform and $M1$ and $M2$ represent modulated polar configurations (Methods). A spin cycloid exists in the ground state of BiFeO$_3$ whereas a complex mixed state becomes increasingly stable in Bi$_{0.85}$La$_{0.15}$FeO$_3$ thin films due to the decreasing energy difference between the two magnetic configurations. \textbf{b} Schematic of the energy landscape of the BiFeO$_3$ and Bi$_{0.85}$La$_{0.15}$FeO$_3$ where the ground state of magnetic textures such as G-type antiferromagnet and spin cycloid phases in the two systems is described. Red arrows form the spin cycloid in the ground state of BiFeO$_3$  with the $\Delta U$ energy barrier whereas the complex state is formed in Bi$_{0.85}$La$_{0.15}$FeO$_3$ due to reduced energy barrier on La substitution. \textbf{c,d}  High angle annular dark field (HAADF) scanning transmission electron microscopy (STEM) images and polar vector mapping in BiFeO$_3$ and Bi$_{0.85}$La$_{0.15}$FeO$_3$. Insets are the schematics of the estimated polarization direction in the unit cell of BiFeO$_3$ and Bi$_{0.85}$La$_{0.15}$FeO$_3$. The average polarization is no longer along $[111]$ after La-substitution. The scale bar is 1 nm. \textbf{e-h} Ferroelectric domain and corresponding magnetic texture of BiFeO$_3$/Bi$_{0.85}$La$_{0.15}$FeO$_3$ in the pristine state. In \textbf{h}, two types of contrast are visible: the stripe-like contrast from the spin cycloid phase, and the more uniform contrast from a canted antiferromagnetic phase.}
\end{figure*}

\lettrine[findent=2pt] Electromagnetic coupling offers a foundational framework for transforming between magnetic and electric fields, primarily facilitated by the principle of magnetic induction through electric currents\upcite{alma991028597229705251,PhysRevLett.6.607}. For applications such as manipulating the magnetization of nanoscale magnets in integrated memory and logic, however, the conventional Oersted field approach has been proven to be energy-inefficient and impractical \cite{Jason_MRAMField}. To address the imperative of low-energy consumption in nonvolatile magnetic memory and logic, a promising new avenue has emerged — direct voltage control of magnetism \cite{chiba2003electrical,ohno2000electric,ohno2015control,spin-drivenferroelectricity,FertRamesh_RMP,yuan2023electrical}. Recent proposals use the magnetoelectric coupling inherent in some multiferroics, which allows for direct electric field control of the magnetic state in such a material\cite{bibes2008towards,spin-drivenferroelectricity}. A notable example of this innovation is the magneto-electric spin-orbit (MESO) device structure, proposed as an inherently non-volatile substitute for complementary metal–oxide–semiconductor (CMOS) devices in integrated logic-in-memory applications \cite{manipatruni2019scalable,manipatruni2018beyond}. To this end, BiFeO$_3$,possessing strong antiferromagnetic magnetoelectric coupling \cite{bibes2008towards,heron2014deterministic,chu2008electric}, is considered a desirable material for MESO-type devices. Additionally, due to antiferromagnetic character, the materials is robust against external magnetic fields and possesses potentially faster-switching dynamics than ferromagnets. Recently, it has also been shown to be an efficient system for demonstrating switchable magnon spin currents \cite{liao2020understanding,Eric_switching}. This electric field switchable electro-magnon coupling allows for a simplified version of the MESO device i.e., the antiferromagnetic state is directly read out using the spin-orbit metal in direct contact with the AFM layer, i.e., without an interleaving ferromagnetic layer. The open question remains: How can we uncover methods to improve performance magnitude and deepen our understanding of spin transport in magnetoelectric multiferroics such as BiFeO$_3$? Addressing these questions has the potential to unlock the application-oriented significance of these materials for broader future problems.

\begin{figure*}[htbp!]
\centering
\includegraphics[width=18cm]{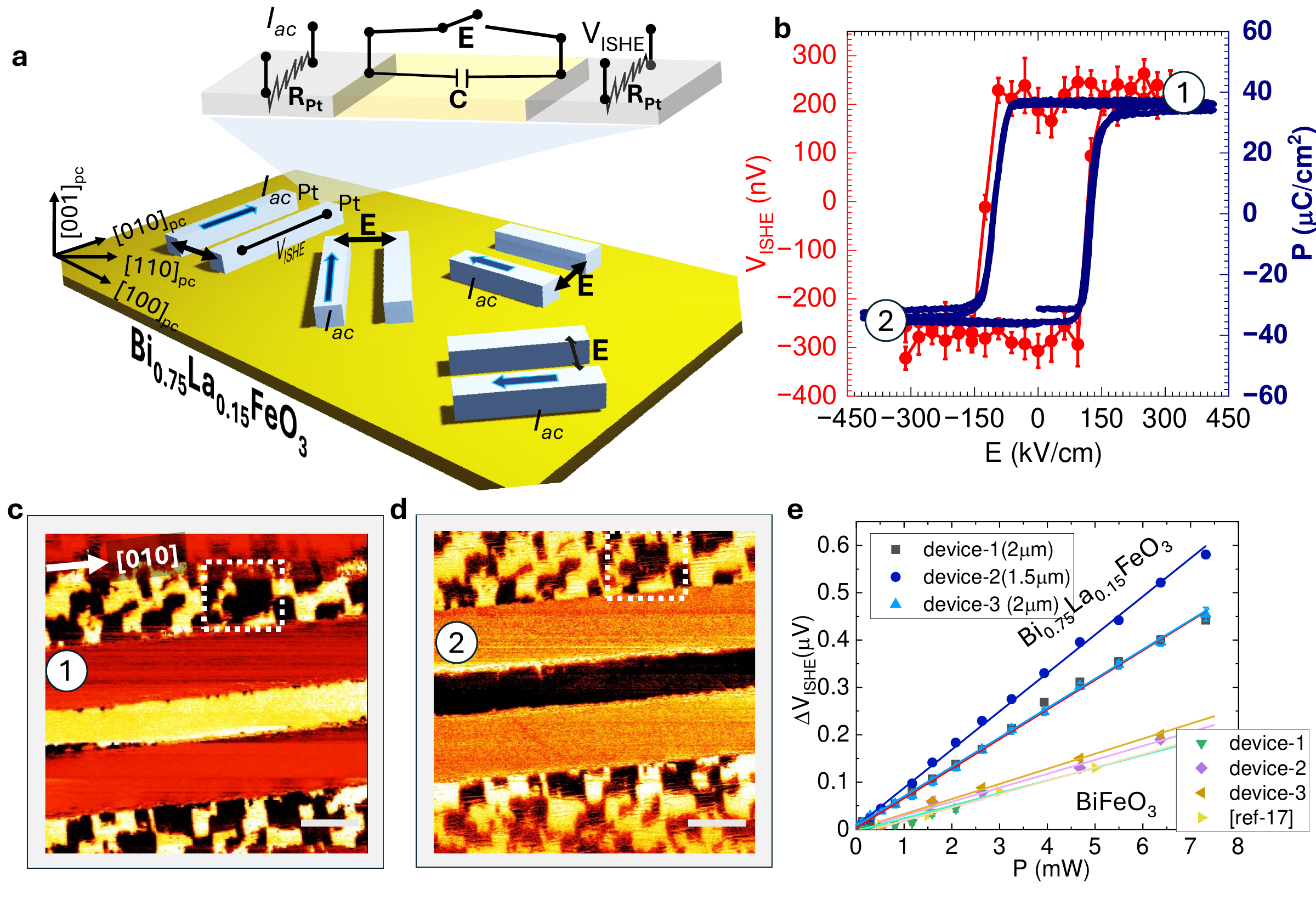}
\caption{\label{fig:magnon1} \textbf{Electric field control of magnons in Bi$_{0.85}$La$_{0.15}$FeO$_3$: a} Nonlocal magnon-transport measurement scheme in Bi$_{0.85}$La$_{0.15}$FeO$_3$ with Pt as a source/detector for spin-charge interconversion via magnon-transport. A resistive circuit schematic of in-plane devices where $R_{Pt}$ is the resistance of Pt electrodes, and $C$ is the capacitance of Bi$_{0.85}$La$_{0.15}$FeO$_3$, where the electric field is mainly distributed. The spin cycloid propagation vector \textit{k} is set by the BLFO polarization, which is controlled by an external in-plane electric field ($E$). The coordinate system uses pseudocubic notations. \textbf{b} Polarization and quasi-static magnon hysteresis as a function of external electric field. The blue line represents the polarization as measured by a Radiant Technologies ferroelectric test system (right axis) and the red circles correspond to the non-local ISHE voltage raw data (left axis). Error bars in ISHE voltage represent the standard statistical variation of lock-in voltages from the least-squares analysis measured over 150s. \textbf{c},\textbf{d} The corresponding PFM images after electrical poling in two opposite directions (labeled by `1' and `2' in \textbf{b}). PFM images were recorded in the same area, as marked by the rectangles. The scale bar is 2$\mu$m. \textbf{e}, Differential voltage ($\Delta$V$_{ISHE}$) recorded in $[010]$ devices as a function of the power injected into the source. Each data point is presented after averaging out to 150s. The Bi$_{0.85}$La$_{0.15}$FeO$_3$ data presented were recorded in several devices with the same orientation and compared with the non-local voltage data belonging to the BiFeO$_3$ (100 nm)/Pt(6nm) with the spacing of 1-2$\mu$m. In the case of BiFeO$_3$, the domains were stripes whereas Bi$_{0.85}$La$_{0.15}$FeO$_3$ data was recorded in a single domain state. Lines are linear fit to the data.}
\end{figure*}

The ground state of bulk BiFeO$_3$ has a large polarization ($\sim$90 $\mu$C/cm$^2$) along $[111]_{pc}$ (pc: pseudocubic) and exhibits a canted G-type antiferromagnetism modulated by a spin cycloid (period $\sim$65 nm due to the inverse spin current effect\cite{sosnowska1982spiral}) below the Néel temperature (640K). BiFeO$_3$ features two principal DMI-like interactions, linked to the polarization and the antiferrodistortive octahedral tilts \cite{jorge_spin_cycloid_PRL}, where the tilts and polarization are strongly coupled \cite{PhysRevB.106.165122,dong2019magnetoelectricity,JorgePRB_MEcoupling}. The octahedral tilt induces a weak magnetic perturbation, and corresponding spin density wave, on top of the antiferromagnetic cycloid of BiFeO$_3$\cite{jorge_spin_cycloid_PRL, JorgePRB_MEcoupling}. This can be imaged directly using scanning Nitrogen-vacancy (NV) magnetometry\cite{manuel2017real}. To introduce tunability in multiferroic properties, rare earth substitution has shown great potential. Often, in these systems, the ferroelectric polarization moves away from $[111]_{pc}$ (hereafter all directions are used in the pseudocubic notation unless otherwise specified) \cite{kan2011chemical,kan2010universal} intoducing competition between ferroelectric and antiferroelectric phases\cite{kan2010universal,yadav2019spatially,yen-lin2020manipulating,husain2024low,prasad2020ultralow}. This may allow for additional switching pathways compared to the parent compound BiFeO$_3$, leading to the possibility for new ferroelectric domain configrations. Understanding the formation of a single-domain multiferroic and its potential as a model system for efficient spin magnon transport is the focus of this work.

Theoretical calculations predict a cycloidal magnetic ground state in BiFeO$_3$, illustrated in Figure \ref{fig:structure}a. La-substitution modifies the structure and impacts both the magnitude and direction of the spontaneous polarization significantly, which is observed to be along $[112]$ and is $\sim$50$\%$ smaller than BiFeO$_3$ . This agrees with experimental values and is supported by high-resolution polar maps (Figure \ref{fig:structure}a,b, Supplementary Note 2). The reduction in spontaneous polarization is accompanied by a corresponding reduction in the polarization dependent DMI interaction strength \cite{Laurent_PRB_DMIvsP} and thus the cycloid becomes less energetically stable. In other words, reducing $P$ enhances the tilting, and consequently, the tilt-induced-canting of the magnetization becomes larger. These findings confirm that La-substitution modifies the energy landscape for both the ferroelectric and antiferromagnetic states in BiFeO$_3$ (Figure \ref{fig:structure} b). In the case of pure BiFeO$_3$, the polar structure is $R3c$ and the cycloid is a stable magnetic state. Interestingly, with the 15$\%$ lanthanum substitution (Methods), the uniform canted moment state ($M1$ and $M2$, Methods) becomes closer in energy to the cycloid state (Figure \ref{fig:structure}a). 

\begin{figure*}[htbp!]
\centering
\includegraphics[width=18cm]{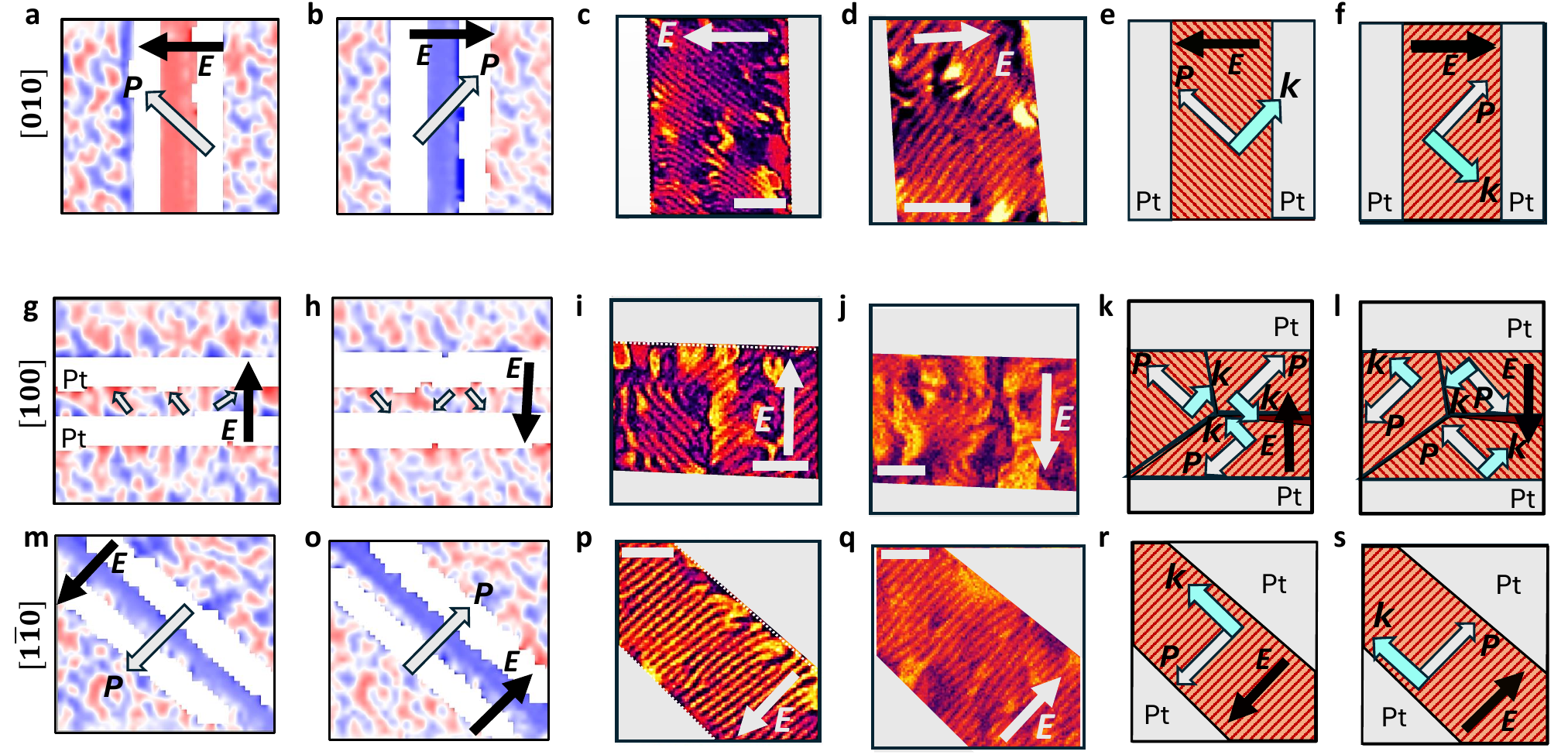}
\caption{\label{fig:NV} \textbf{Electric field control of magnonics based on the controllable magnetic and polar texture:} \textbf{a-f}, \textbf{g-l}, and \textbf{m-s} SHG-Linear dichroism maps, NV magnetometry images, and corresponding schematic for devices $[010], [100]$ and $[\bar{1}10]$. SHG (for in-plane polarization) and NV (for spin magnetic texture) are recorded for two opposite poling directions. Dark arrows represent the direction of the electric field and gray arrows show the polarization direction in specific domains. The gray pads are used for the application of an in-plane electric field. The arrow indicates the direction of the applied electric field. The stripe patterns in NV images are indicative of the canonical spin cycloid as observed in previous studies \cite{meisenheimer2023persistent,manuel2017real,meisenheimer2024designed}. Schematics highlight the polarization $P$ and spin cycloid propagation direction $k$ deduced from NV measurements. The scale bar in NV images is 500nm.}
\end{figure*}

In this spirit, Bi$_{0.85}$La$_{0.15}$FeO$_3$ films have been deposited on $(110)_O$ (O: orthorhombic) DyScO$_3$ substrates (Methods)(Extended Data Fig. \ref{fig:structure}, and Supplementary Note 1, 2).  Using piezo-force microscopy (PFM) and NV magnetometry, as predicted (Figure \ref{fig:structure} a-b), the pure cycloid (within 71$\degree$ ferroelectric BiFeO$_3$) and mixed state of cycloid+G-type antiferromagnet (in blocky-mixed ferroelectric Bi$_{0.85}$La$_{0.15}$FeO$_3$) are both observed in a mixed equilibrium state (Figure \ref{fig:structure} e-h). 
To then understand the effect of electric field on the as-grown ferroelectric domain structure, and therefore the ferroelectric polarization, in-plane capacitors were fabricated by optical lithography (\textit{ex-situ} sputtered platinum (Pt) wires 120$\mu$m $\times$ 1.3$\mu$m $\times$ 15nm, with $\sim$2$\mu$m spacing and resistivity of $\sim$20 $\mu\Omega$ cm). The devices were patterned along four different angles in which the long-axis of Pt electrode pairs are parallel to the substrate $[100]$, $[010]$, $[110]$, and $[\Bar{1}10]$ pseudocubic directions (Figure \ref{fig:magnon1} a). To visualize the ferroelectric domain reversal across the in-plane devices (Figure \ref{fig:magnon1} b, $P$ vs $E$ hysteresis), PFM images were recorded after poling in two opposite electric field directions (Fig. \ref{fig:magnon1}(c) and (d) and Fig. \ref{fig:NV}). For a field applied along the $[100]$ direction, in-plane poling leads to the formation of a single ferroelectric domain, which is the novel feature of Bi$_{0.85}$La$_{0.15}$FeO$_3$. This has a powerful impact on the magnetic cycloid, which is particularly important for spin transport (discussed later). The formation of a single ferroelectric domain is further verified by rotating the device and performing PFM imaging (Supplementary Note 3), which shows the uniform domain contrast indicative of a single ferroelectric domain. Previously, monodomain features were realized through a non-trivial approach in BiFeO$_3$ using a scanning probe-tip-based method in slow scan mode to physically write a monodomain using a localized in-plane electric field from the tip \cite{manuel2017real,balke2009deterministic}, requiring time and an extremely careful experimental protocol \cite{matzen2014super,Ramesh_localmonodomainPRL}, compared to the direct voltage pulse induced switching of La substituted BiFeO$_3$ single domain that we have adopted in this work. 

In the case of Bi$_{0.85}$La$_{0.15}$FeO$_3$, the polarization is deterministically switched by an electric field at the macroscopic scale of hundreds of microns (see Extended Data Figure 3). A key result of this study is the fact that switching the polarization state with a single, in-plane pulse leads to the deterministic switching and formation of a single multiferroic domain (details in Supplementary Notes 3-5). However, for a field applied along the $[010]$ direction, that is, Pt wires parallel to $[100]$, a blocky multidomain case persists even in the poled region (Supplementary Figure 6-8). In this multidomain case, upon poling, the domains are locally switched (Supplementary Figure 8), where the domain wall boundaries (or antiphase boundaries, Supplementary Figure 9) do not move. This asymmetric behavior can be attributed to the anisotropic strain from the substrate (Supplementary Note 1), preventing the formation of a macroscopic domain in the device $[100]$. In devices with electrodes parallel to $[1\Bar{1}0]$ and $[110]$, a single ferroelectric domain is formed which can be expected since a component of the electric field points along $[100]$, allowing the antiphase boundaries to nucleate and move with the field.
We can now use such a single domain multiferroic as a model system to understand the stability of the spin cycloid and the corresponding   spin transport.

\begin{figure*}[htbp!]
\centering
\includegraphics[width=17cm]{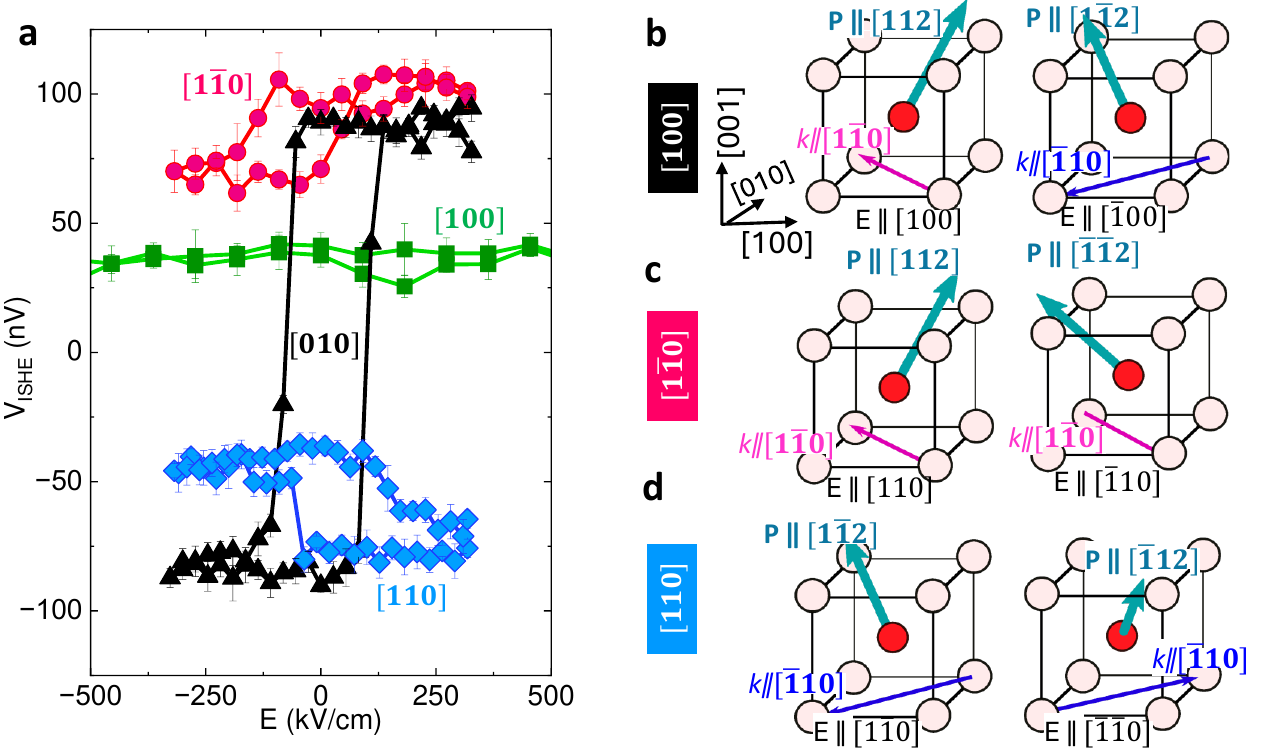}
\caption{\label{fig:magnon2} \textbf{Anisotropic magnon transport:} \textbf{a}
Magnon-generated nonlocal ISHE voltage hysteresis measured as a function of the external in-plane electric field in devices with four different orientations labeled by the pseudocubic direction of the Pt wires. The power in the source electrode was fixed to 2mW ($I_{ac}$=1.7mA). A depiction of the response of the Bi$_{0.85}$La$_{0.15}$FeO$_3$ unit cell to different poling fields is given in \textbf{b-d} for device orientations $[010]$, [$\bar{1}$10], and $[110]$ respectively. $k$, $E$, $P_{net}$ represent the propagation vector of the spin cycloid, in-plane electric field, and net in-plane ferroelectric polarization. The relation between $P$ and $k$ is drawn based on the experiment performed in Figure~\ref{fig:NV} where $P$ and $k$ are mapped out, and are consistent with prior studies of the cycloid in BiFeO$_3$ \cite{meisenheimer2023persistent,manuel2017real,meisenheimer2024designed}. The schematics are only considered here for single-domain devices, and for the multidomain device [100] where the ISHE voltage change as a function of the electric field is found to be negligible (green data in \textbf{a}), more discussion can be found in Supplementary Note 6.}
\end{figure*}

To probe the spin transport, first, an in-plane electric field was applied between the source and detector wires, as indicated in Figure \ref{fig:magnon1}a. Following each electrical pulse, a low-frequency (7Hz) alternating current is introduced into the source wire, generating a magnon spin current through the spin Seebeck effect. Subsequently, a non-equilibrium magnon spin accumulation at the Bi$_{0.85}$La$_{0.15}$FeO$_3$ interface underneath the Pt detector initiates the flow of spin angular momentum into the adjacent Pt. The resulting spin current is then converted into a measurable voltage through the Inverse Spin Hall Effect (ISHE) of Pt (Supplementary Note 6), and the signal is lock-in detected at 2$\omega$. Each data point is averaged over a duration of 150 seconds. The ferroelectric polarization hysteresis was measured at 5kHz (Methods) and the corresponding ISHE hysteresis was recorded in a remnant state where an electric field was applied only to set the polarization state and removed during the nonlocal voltage (V$_{ISHE}$) measurement. The nonlocal voltage hysteresis precisely reflects the ferroelectric polarization response (Figure \ref{fig:magnon1}b, red data), indicating the existence of polarization-controlled magnon transport. Notably, in the $[010]$ devices, the electric field and therefore the polarization $P$ has the capacity to control the sign of the magnon spin current flowing through the Bi$_{0.85}$La$_{0.15}$FeO$_3$. This nonvolatile electric field magnon switching is illustrated in the extended data Figure 4, where the ferroelectric polarization deterministically controls non-reciprocal magnon transport in the Bi$_{0.85}$La$_{0.15}$FeO$_3$.

Similar experiments on BiFeO$_3$  with a stripe domain structure were performed and a comparison is presented in Fig. \ref{fig:magnon1}e. The data corresponding to BiFeO$_3$  is also reproduced from Parsonnet \textit{et al} \cite{Eric_NLBFO}. The data from the different devices corresponds to the 71$\degree$ BiFeO$_3$ and reported data from Parsonnet \textit{et al} \cite{Eric_NLBFO} belongs to the 109$\degree$ BiFeO$_3$. We find that the Bi$_{0.85}$La$_{0.15}$FeO$_3$ has a consistently higher voltage output than the BiFeO$_3$ (by $\sim$400$\%$ at the equivalent spacing). Furthermore, we find that the magnitude of the electric field required to switch the magnon spin current is indeed significantly smaller (Supplementary Figure 23), consistent with prior studies \cite{yen-lin2020manipulating}. This doubly confirms the key advantages of single-domain Bi$_{0.85}$La$_{0.15}$FeO$_3$ over its parent compound.

The strong enhancement in the inverse spin Hall voltage for the Bi$_{0.85}$La$_{0.15}$FeO$_3$ compared to BiFeO$_3$ prompts us to explore the microscopic differences, if any, in the magnetic structure, particularly the spin cycloid.  We used a combination of imaging techniques (PFM and SHG-linear dichroism to probe the ferroelectric state and NV magnetometry to probe the spin cycloid, details in Methods).  A comparison of the ferroelectric domain structure and corresponding magnetic (spin cycloid) is presented in Figure \ref{fig:NV}. To determine the local directions of the polarization in each domain (discussed in Figure \ref{fig:magnon1} c,d), optical SHG is used to map the ferroelectric domains in oppositely poled states (Figure \ref{fig:NV} a,b). The red and blue areas correspond to domains with orthogonal in-plane polarization, and it is clear that in the device $[010]$, the in-plane polarization is switched by 90$\degree$ upon poling with oppositely directed fields. NV microscopy (Fig. \ref{fig:NV}c,d) reveals the presence of uniform spin cycloids in oppositely poled domain. It is noteworthy that the sense of the cycloid stripes has rotated by 90 degrees, between these two switched states. This observation reveals that the ferroelectric single domains prefer to form a single variant cycloid, consistent with previous results \cite{manuel2017real}. We can conclude that the polarization is parallel to the spin cycloid stripes, which leads us to conclude that $P$ is orthogonal to the propagation vector $k$ (drawn schematically in Figure \ref{fig:NV}e,f ), a result that is consistent with previous works \cite{manuel2017real, meisenheimer2023persistent,meisenheimer2024designed}. 
It is also validated by poling the Bi$_{0.85}$La$_{0.15}$FeO$_3$ devices at different angles with respect to the direction of the spin cycloid and the ferroelectric polarization.  The multi-domain device $[100]$ has two variants of cycloid corresponding to the two ferroelectric domains (Figure \ref{fig:NV}g-l), whereas the same contrast in $[1\bar{1}0]$ (in opposite poling) indicates 180$\degree$ switching (Figure \ref{fig:NV} m-s). Despite the same $k$ cycloid in 180$\degree$ switch, $P$ switching will change the handedness in the opposite poled state \cite{Laurent_PRL_cycloidswitchwithP}. With these insights, we conclude that the magnetic and ferroelectric order parameters are intimately tied in Bi$_{0.85}$La$_{0.15}$FeO$_3$ in a similar fashion to BiFeO$_3$ \cite{manuel2017real,meisenheimer2023persistent}, and we show how the polarization and cycloid behave under electric fields pointing in different directions.

To probe the effect of such a single variant spin cycloid in the single-domain ferroelectric state, we proceeded to measure the non-local spin transport through the same test structures described in Figure \ref{fig:NV} a-s, electric field dependent inverse spin Hall voltage hystereses were measured along these crystallographic directions under the same protocol as discussed in Figure \ref{fig:magnon1}a,b. The single-domain devices $[010]$ show ISHE voltage hysteresis  (in BLACK) that corresponds to their ferroelectric hysteresis (Figure \ref{fig:magnon1}b). Strikingly however, the multi-domain $[100]$ oriented device does not show any appreciable ISHE hysteresis (in GREEN) despite exhibiting a clear ferroelectric hysteresis (Supplementary Figure 7).  Insight into this is readily obtained from the NV magnetometry images shown in Figure \ref{fig:NV} i,j which shows no change in the topology of the spin cycloid; this is also schematically captured in Figure \ref{fig:NV} k,l. This reveals that not only is magneto-electric coupling important but also the uniform magnetic texture is required for effective magnon spin flow. The behavior exhibited by the $[010]$ device serves as a key to understanding the $[100]$ device's behavior. A spin cycloid propagation vector of $k=[1\bar{1}0]$ results in a positive ISHE signal, and $k=[\Bar{1}10]$ yields a negative ISHE signal, as shown in Fig. \ref{fig:magnon2}a; it follows that a combination of domains with $k=[\Bar{1}10]$ and $k=[1\bar{1}0]$, as observed in the $[100]$ device, leads to a null signal without any discernible magnon spin hysteresis. Although the precise correlation between the direction of $k$ and the spin carried by a magnon current would be interesting, the present observations affirm that the direction of $k$ holds greater significance than the net polarization in determining the non-local magnon signal.

Within the $[1\Bar{1}0]$ device, illustrated in Figure \ref{fig:magnon2}, we note that the  $[1\Bar{1}0]$ ($[110]$) devices have a lower magnitude with a positive (or negative) offset. The sign of the offset is consistent from device to device (5 devices for each orientation), as discussed in Extended Data Figure 5. This can be understood from the symmetry of $P$ and $k$, if we consider that La-substitution can allow for different symmetry operations when switching the polarization\cite{meisenheimer2023persistent,meisenheimer2024designed}. With an electric field along $[110]$, in the parent BiFeO$_3$, this would result in a C$_2$ rotation about $[001]$, or two successive 71$\degree$ switches within the $(001)$ \ref{fig:magnon2}b. Here, $k$ is also rotated about the $[001]$, to which it is orthogonal, resulting in $k \rightarrow -k$. In the case of La-substituted BFO, however, the polarization along $[112]$ may allow for this rotation to happen about the $[1\bar10]$ or $[\bar{1}10]$ direction, rather than about the film normal (Figure \ref{fig:magnon2}c,d). This operation, for example from $[112]$ to $[\bar1\bar12]$ is only a $\sim$70$degree$ rotation of $P$, rather than the two 71$\degree$ events to rotate around $[001]$. Additionally, the rotation axis in this scheme is parallel to $k$, which then does not change sense after the operation. We would expect this to result into a small magnon signal, as observed. This elucidates the anisotropic nature of magnon transport as it is intricately linked to the spin cycloid and thus the polarization of the Bi$_{0.85}$La$_{0.15}$FeO$_3$.

In summary, our study demonstrates the effective transmission of magnons in lanthanum-substituted BiFeO$_3$, resulting in a multiferroic material that can be polarized into a stable non-volatile uniform ferroelectric domain with a single variant of the spin cycloid. This stands in contrast to pure BiFeO$_3$, where the coexistence of two variants in both spin cycloids and stripe-like ferroelectric domains leads to a diminishing magnon signal. We observe that -- by suitably choosing the direction of the applied electric field -- it is possible to maximize or cancel the effect of ferroelectric switching on magnon transport. This research provides a means to customize ferroelectric domains and complex antiferromagnetic spin cycloids, as well as to understand the resulting spin transport, offering a pathway to design the single domain multiferroics for efficient magnon transport for future applications.

\hfill \break
\noindent\textbf{\Large{Methods}}

\noindent\textbf{Thin film deposition}\\
BiFeO$_3$ and Lanthanum (La) substituted BiFeO$_3$ (Bi$_{0.85}$La$_{0.15}$FeO$_3$) thin films were prepared by pulsed laser deposition (PLD) in an on-axis geometry with a target-to-substrate distance of $\sim$50 mm using a KrF excimer laser (wavelength 248 nm, COMPex-Pro, Coherent) on DyScO$_3$ $(110) $substrates. Film thickness was fixed to 90 nm unless otherwise specified. Before the deposition, the substrates were cleaned with IPA and Acetone for 5 min each. The substrates were attached to a heater using silver paint for good thermal contact. 
BiFeO$_3$ and Bi$_{0.85}$La$_{0.15}$FeO$_3$ layers were deposited with a laser fluence of 1.8 Jcm$^{-2}$ under a dynamic oxygen pressure of 140 mTorr at 710 $\degree$C with a 15 Hz laser pulse repetition rate. The samples were cooled down to room temperature at 30 $\degree$C/min at a static O$_2$ atmospheric pressure. The prepared samples were immediately transferred to a high vacuum DC magnetron sputtering chamber for Pt deposition. 15 nm of Pt was sputtered at 15W power at room temperature in a 7 mTorr dynamic Ar atmosphere. The thicknesses were calibrated using X-ray reflectivity and atomic force microscopy.
\\

\noindent\textbf{Crystal Structure Determination}\\ The crystal structures of both BiFeO$_3$ and La-substituted BiFeO$_3$ were determined through X-ray diffraction, utilizing a high-resolution X-ray diffractometer (PANalytical, X’Pert MRD). The symmetric line scan ($\theta$–2$\theta$) employed a fixed-incident-optics slit set at 1/2$\degree$, while the reciprocal space mapping (RSM) involved an asymmetric 2D scan with a slit of 1/32$\degree$. The X-ray source was used the Cu K$\alpha$ transition (wavelength: 1.5401 \AA), and detection employed a PIXcel$^{3D}$-Medipix$^3$ detector with a fixed receiving slit of 0.275 mm.\\

\noindent\textbf{Cross-section Sample Preparation and High-angle Annular Dark Field Scanning Transmission electron microscopy (HAADF-STEM):}\\
The cross-section samples were prepared using a Helios660 scanning electron microscope/focused ion beam (SEM/FIB) with a gallium (Ga) ion beam source. After sample preparation, the cross-section samples were analyzed using an FEI Titan Themis G3 scanning transmission electron microscope (STEM) equipped with double correctors and a monochromator. High-angle annular dark-field scanning transmission electron microscopy (HAADF-STEM) imaging was performed at 300 kV accelerating voltage.
Fourier-filtered HAADF-STEM images were analyzed using CalAtom software to extract the atomic position of Bi/La and Fe ions by multiple-ellipse fitting. The Fe displacement vector in each unit cell was calculated by confirming the center of mass of its four closest Bi/La neighbors. The displacement vector D of the Fe column is represented as follows:
\begin{equation}
    \textbf{D}=\textbf{r}_{Fe}-\frac{\textbf{r}_1+ \textbf{r}_2+\textbf{r}_3+\textbf{r}_4}{4},
\end{equation}
where $\textbf{r}_{Fe}$ is the position vector of the Fe column. $\textbf{r}_{1}, \textbf{r}_{2}, \textbf{r}_{3}, \textbf{r}_{4}$ are the position vectors of the four closest Bi/La neighbors in each unit cell. The color of the displacement vectors was represented by the vector magnitude.  The visualization of the two-dimensional atomic displacement was carried out using Python. Calculation of the net displacement according within the unit cell projection is discussed in the supplementary.

\noindent\textbf{Ferroelectric Domain Characteristics}\\
Piezoresponse force microscopy (PFM) imaging was conducted employing the MFP-3D system from Asylum Research, featuring Dual AC Resonance Tracking (DART) mode. Throughout the imaging process, the system operated in lateral mode, ensuring accurate lateral resolution in the acquired images. For these measurements, a silicon cantilever coated with platinum (Pt) was utilized, serving as a conducting electrode for the precise and localized application of an electric field. See Supplementary Note 2 for further information.\\

\noindent\textbf{Optical second harmonic generation for in-plane polarization mapping (SHG)}\\
These measurements were conducted in a normal-incidence reflection geometry on poled devices. Light excitation was achieved using a Ti/sapphire oscillator with $\sim$ 100 fs pulses, a center wavelength of 900 nm, and a 78 MHz repetition rate. To manipulate the incoming light's polarization, a Glan–Thompson polarizer was employed, followed by passage through a half-wave plate. The polarized light then traversed a short-pass dichroic mirror and was focused onto the sample using a 100x objective lens with a numerical aperture (NA) of 0.95. The back-reflected SHG signal passed through a short-pass filter and was detected using a spectrometer (SpectraPro 500i, Princeton Instruments) equipped with a charge-coupled device camera (Peltier-cooled CCD, ProEM+:1600 eXcelon$^3$, Princeton Instruments). Diffraction-limited confocal scanning microscopy was employed to generate SHG intensity maps. At the sample location, a commercial Thorlabs polarimeter verified the incoming light's polarization incident on the sample and the light polarization entering the detector. Linear dichroism maps were constructed through the subtraction of SHG intensity maps with incident light polarization along $[110]_{pc}$ or $[1\bar{1}0]_{pc}$ directions. The poling process was performed \textit{ex-situ} for all devices. See Supplementary Note 3 for further information.\\

\noindent\textbf{Scanning Nitrogen-Vacancy (NV) microscopy}\\
The magnetic texture in the samples was imaged at room temperature utilizing a commercial scanning NV magnetometer (Qnami ProteusQ). Scanning NV magnetometry has been described extensively elsewhere; briefly a parabolically-tapered diamond cantilever (Quantilevel MX+) was used to detect the stray fields from the sample by probing the frequency shift of the NV center spin as the tip was scanned across the surface. To facilitate wide-area scans, data was collected in the "iso-B" mode, where the peak shift is estimated from the microwave response at two frequencies rather than the full spectrum (e.g., Ref. \cite{tetienne2014nanoscale}). Iso-B measurements were validated against select measurements of the full spectrum to ensure the magnetic texture is reported faithfully (See Supplementary Note 5)

\noindent{\textbf{Device Fabrication}}\\ 
The sample fabrication started with sonication in acetone and isopropyl alcohol. Subsequently, a positive photoresist (MIR 701), approximately 500 nm thick, was uniformly coated at 7000 RPM for 60 seconds using a spin coater. The coated sample was then baked at 100 $\degree$C for 60 seconds. Photolithography was executed through a Karl Suss MA6 Mask Aligner, with i-line exposure at 10 mW/cm$^2$ for 5 seconds. Following exposure, the resist underwent wet-etching using MEGAPOSIT MF-26A photoresist developer for 20 seconds. Subsequently, the Pt layer was ion-milled down to the multiferroic film surface (Intlvac Nanoquest, with a Hiden Analytical SIMS), resulting in the formation of rectangular stripes measuring 120 $\mu$m × 1.3 $\mu$m. This process was conducted at the Marvell Nanofabrication laboratory at UC Berkeley.\\

\noindent\textbf{Spin Transport Measurements}\\
Transport measurements were conducted employing 4-terminal devices, wherein two terminals were dedicated to source current injection, and the remaining two served as output terminals for inverse spin Hall effect (ISHE) voltage measurement. One source terminal and one detection terminal were also used to apply an electric field for ferroelectric polarization control. The entire experimental setup and procedures were orchestrated using an in-house developed Python code and a Keithley 7001 switch box, maximizing repeatability. 
To measure the nonlocal ISHE voltage ($V_{ISHE}$), an SR830 lock-in amplifier was synchronized to the second harmonic of the 7Hz source current, isolating responses to the thermal gradients. This comprehensive setup allowed us to perform accurate and controlled transport measurements (using all automated codes), facilitating the investigation of electric field-controlled nonlocal voltage measurements. See Supplementary Note 6 for more information.\\

\noindent\textbf{Computational Methods (Effective Hamiltonian):}\\
In the case of BiFeO$_3$, the magnetic ground state is a G-type antiferromagnetic configuration, which is modulated by the complex magnetic arrangement called a spin cycloid. The BiFeO$_3$ doped with rare-earth leads to further modulation in the magnetic texture or relaxed into a G-type configuration without the cycloid. To understand this complex state in BiFeO$_3$ and doped BiFeO$_3$ compounds, we performed Monte Carlo simulations governed by the first principle-based effective Hamiltonian. This effective Hamiltonian is expressed as follows for BiFeO$_3$ and doped BFO:
\begin{equation}\label{BFO_Hamiltonian}
\begin{split}
E_{\text{total}}=E_{\text{{FE}-{AFD}}}(\{ \mathbf{u}_i\}, \{ \mathbf{\omega}_i\}, \{ \eta_{\text{H}}\}, \{ v_i\})\\
+E_{\text{mag}} (\{ \mathbf{m}_i\}, \{ \mathbf{u}_i\}, \{ \mathbf{\omega}_i\}, \{ \eta_{\text{H}}\})\\,
\end{split}
\end{equation}
where the first term in equation (2) $E_{\text{{FE}-{AFD}}}$ (FE: ferroelectric, AFD: antiferrodistortion octahedral tilts) contains energy terms arising from the nonmagnetic variables (local mode ($\mathbf{u_i}$) being the parameter corresponding to the electric dipole (or the electrical polarization), global homogeneous ($\eta_{\text{H}}$) and Fe-centred inhomogeneous strain tensor ($v_i$).  $\mathbf{\omega}_i$ is the oxygen octahedral tilt representing the axis of rotation) and their couplings. The second term represents the magnetic mode of the BiFeO$_3$ ($m_i$ represents the magnetic moment at site $i$ centered at the Fe ion with its magnitude fixed (4$\mu_B$)) and its couplings with other modes. The expansion of this term is as follows:
\begin{equation}\label{BFO_Hamiltonian}
\begin{split}
E_{\text{mag}} (\{ \mathbf{m}_i\}, \{ u_i\}, \{ \mathbf{\omega}_i\}, \{ \eta_i\})\\
= \sum_{i,j,\alpha,\gamma} Q_{ij \alpha \gamma} m_{i\alpha} m_{j\gamma}+
\sum_{i,j,\alpha,\gamma} D_{i j \alpha \gamma} m_{i\alpha} m_{j\gamma}\\
+\sum_{i,j,\alpha,\gamma, \nu, \delta} E_{ij\alpha \gamma \nu \delta} m_{i\alpha} m_{j\gamma} u_{i\nu} u_{i\delta}\\
+\sum_{i,j,\alpha,\gamma, \nu, \delta} F_{ij\alpha \gamma \nu \delta} m_{i\alpha} m_{j\gamma} \omega_{i\nu} \omega_{i\delta}
\\
+\sum_{i,j,l,\alpha, \gamma} G_{ij\alpha\gamma} \eta_l (i) m_{i\alpha} m_{j\gamma}\\
+\sum_{i,j} K_{i j} (\mathbf{\omega}_i-\mathbf{\omega}_j)\cdot (\mathbf{m}_i\times \mathbf{m}_j)\\
+\sum_{i,j} C_{i j} (\mathbf{u}_i\times \hat{e}_{i,j})\cdot (\mathbf{m}_i\times \mathbf{m}_j).
\end{split}
\end{equation}\\
Here the I$^{st}$ term represents the magnetic dipolar interaction. The II$^{nd}$  term corresponds to the magnetic exchange coupling up to the third nearest neighbor. The III$^{rd}$, IV$^{th}$, and V$^{th}$ terms describe the change in the magnetic exchange interaction induced by the local polar mode, AFD tilt, and strain. An important point to note is that the first five energy terms lead to the collinear magnetism in BiFeO$_3$. The VI$^{th}$ term involving octahedral or AFD tilting represents the  Dzyaloshinskii–Moriya interaction (DMI) and is responsible for the weak magnetization in the AFM state of BiFeO$_3$. The last term of Eq. (4) is responsible for the cycloid (via the inverse spin-current effect which is a DMI effect), and it is the only term related to electric polarization. This energy allows the stable spin cycloid with $k$ being the propagation vector along $[1\bar{1}0]$ (within (111) plane) (with P $\parallel$ $[111]$) in BiFeO$_3$.
All the coupling coefficients were calculated using Density Functional Theory for both pure BiFeO$_3$ as well as lanthanum-doped BiFeO$_3$. All the calculations were done for bulk stress-free supercells of $12\times12\times12$ unit-cells, both for pure and doped BiFeO$_3$. The complex modulated phases $M1, M2$ are phases found as a result of temperature cooling of rare-earth-doped BiFeO$_3$, further relaxed for (15$\%$) La-substituted BiFeO$_3$ and represent modulated polar arrangements of periods of 6 and 4 unit cells respectively. 
\\

\noindent\textbf{DATA AVAILABILITY}\\
The data that support the findings of this study are available from the corresponding author upon reasonable request.\\

\bibliography{apssamp}

\hfill \break
\noindent\textbf{\large{Acknowledgements}}\\
We thank Sasikanth Manipatruni, and Dmitri E. Nikonov for the fruitful discussions. This work was primarily supported by the U.S. Department of Energy, Office of Science, Office of Basic Energy Sciences, Materials Sciences and Engineering Division under Contract No. DE-AC02-05-CH11231 (Codesign of Ultra-Low-Voltage Beyond CMOS Microelectronics) for the development of materials for low-power microelectronics. H.Z. and R.R. acknowledge the Air Force Office of Scientific Research 2D Materials and Devices Research program through Clarkson Aerospace Corp under Grant No. FA9550-21-1-0460. P.K. acknowledges support from the Intel Corporation as part of the COFEEE program. S.Z. and L.C. acknowledge funding from Brown School of Engineering and Office of the Provost. P.S. acknowledges support from the Massachusetts Technology Collaborative, Award number 22032. L.W.M. and R.R. also acknowledge partial support from the Army/ARL as part of the Collaborative for Hierarchical Agile and Responsive Materials (CHARM) under cooperative agreement W911NF-19-2-0119. X.L. and Y.H. are supported by the Welch Foundation (C-2065-20210327). Y. H. acknowledges the support from NSF-2329111 and NSF-2239545). We acknowledge the Electron Microscopy Center, Rice. J.I.G. acknowledges support from the Luxembourg National Research Fund through grant C21/MS/15799044/FERRODYNAMICS. B.X. acknowledges financial support from the National Natural Science Foundation of China under Grant No. 12074277. S.M. and L.B. would like to acknowledge the ARO Grant No. W911NF-21-1-0113, the U.S. Department of Defense under the DEPSCoR program (Award No. FA9550-23-1-0500) and the Vannevar-Bush Faculty Fellowship (VBFF, Grant No. N00014-20-1-2834). The calculations were performed at the Arkansas High Performance Computing Center (AHPCC).

\hfill \break
\noindent\textbf{\large{Author contributions}}\\
R.R. and S.H. conceived the idea and designed the experiments. S.H. and I.H. performed sample growth and measurements. I.H. patterned the devices and wrote the experimental procedure scripts. P.M. and P.S. performed the Nitrogen-Vacancy (NV) magnetometry. S.M. performed theoretical calculations under the supervision of B.X., J.I., and L.B. X.L. performed cross-sectional microscopy and polarization mapping under the guidance of Y.H. M.R. did controlled sample preparation under the supervision of D.S. and performed some NV measurements with the help of L.C. P.B. performed second harmonic generation (SHG) mapping. H.T. performed sample lithography under the supervision of J.G.A. J.K., P.K., T.Y.K., and H.Z. gave suggestions on the PFM experiments. S.S. and L.W.M. gave suggestions and commented on the manuscript. R.R. and Z.Y. supervised the work. S.H. and I.H. wrote the manuscript draft. All authors have participated in the discussion and reviewed the manuscript.

\hfill \break
\noindent\textbf{\large{Competing interests}}\\
The authors declare no competing interests\\

\clearpage
\newpage

\onecolumngrid

\section*{Extended Data}
Extended data for "Efficient magnon transport in a single domain multiferroic"

\setcounter{figure}{0}
\renewcommand{\figurename}{\textbf{Extended Data Figure}}

\begin{figure*}[htbp!]
\centering
\doublespacing
\includegraphics[width=17cm]{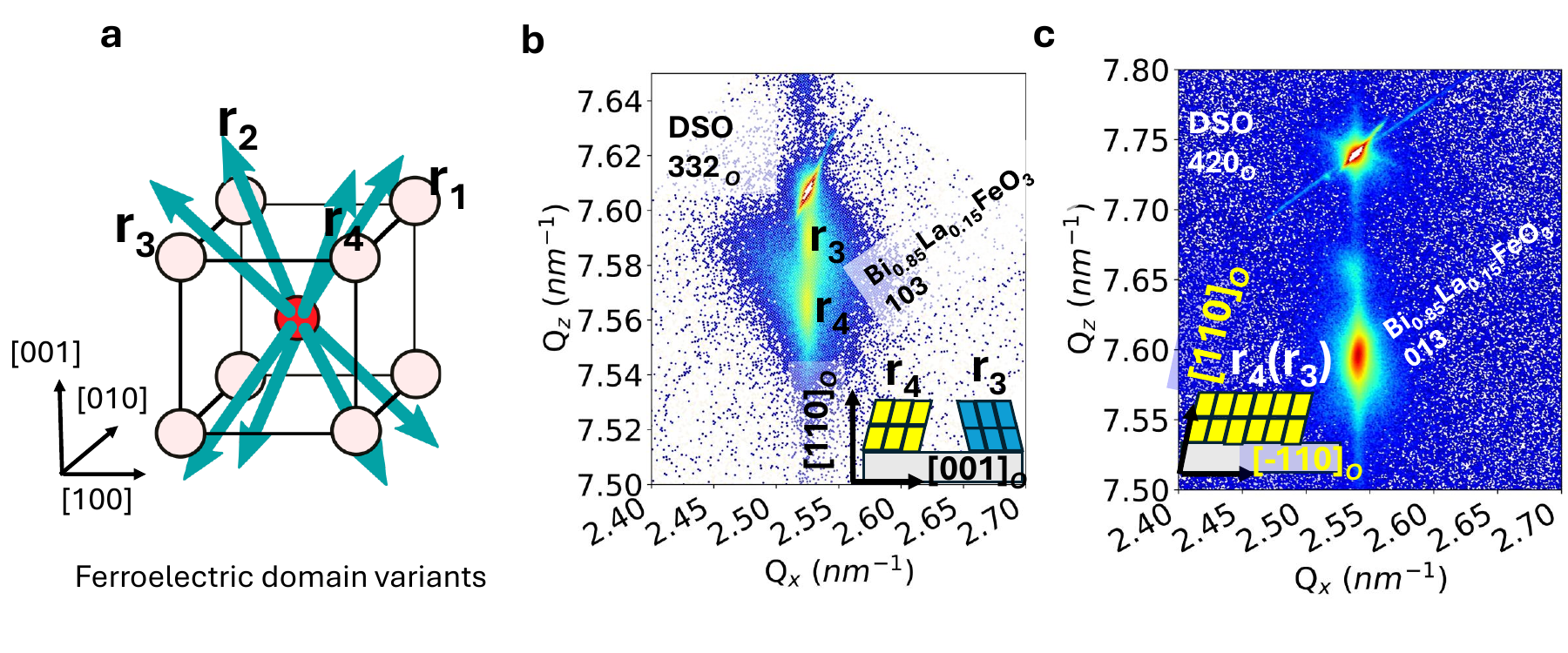}
\caption{\label{fig:RSM} \textbf{Reciprocal mapping of Bi$_{0.85}$La$_{0.15}$FeO$_3$:} a Schematic of Ferroelectric domain variant r$_1$, r$_2$, r$_3$, r$_4$. Asymmetric reciprocal space maps were recorded on Bi$_{0.85}$La$_{0.15}$FeO$_3$ thin films of 90 nm. Reciprocal space mapping of (332)$_O$ planes in DSO shows the $(103)$ planes in Bi$_{0.85}$La$_{0.15}$FeO$_3$, suggesting that the Bi$_{0.85}$La$_{0.15}$FeO$_3$ is epitaxially strained to the DSO substrate. Using RSM, the in-plane (out-of-plane) lattice parameters of BLFO were found to be 3.960 (3.965 Å). The peak splitting occurs in the \textit{h0l}-diffraction condition along $(103)$, but not in the \textit{0kl}-diffraction condition $(013)$. This indicates that only two structural variants are present in these films grown on (220) DyScO$_3$ substrates, consistent with the PFM results in previous studies \cite{yen-lin2020manipulating}. The two peaks in 103 reflections represent the $R$-like phase with the $M_{A}$ structure, which is in agreement with previous reports \cite{chu2007ferroelectric}. A schematic illustration of the two-domain motif, as viewed along the [1$\Bar{1}$0]$_O$ and [001]$_O$. Because the substrate has a monoclinic distortion along the [0$\Bar{1}$1] ([010]$_O$), only two structural variants $r_3$ and $r_4$, which have a spontaneous shear distortion along [$\Bar{1}$$\Bar{1}$1] and [$\Bar{1}$1], respectively give rise to a net shear distortion along the monoclinic distortion of the substrate. Therefore, these two variants $r_3$ and $r_4$ are energetically favorable when the rhombohedral films are grown on $[110]_O$ substrates, in order to follow the substrate monoclinic distortion.}
\end{figure*}

\clearpage
\newpage

\begin{figure*}[htbp!]
\centering
\doublespacing
\includegraphics[width=17cm]{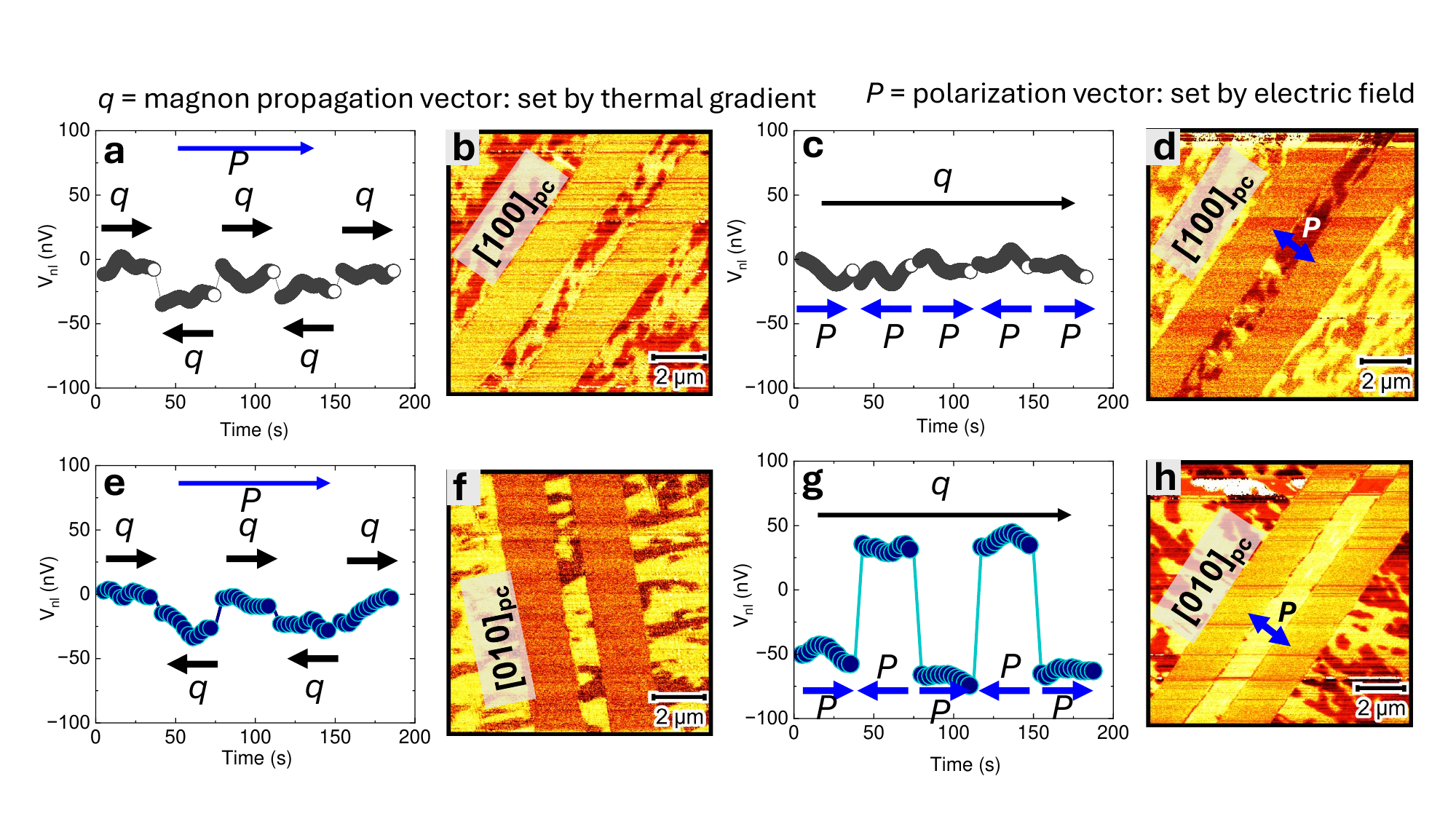}
\caption{\label{fig:RSM} \textbf{Emerging magnon transmitted voltage signal in pristine and poled Bi$_{0.85}$La$_{0.15}$FeO$_3$: }\textbf{a-d} and \textbf{e-f} are the results on $[100]$ and $[010]$ devices, respectively. \textbf{a}, Nonlocal inverse spin Hall voltage (V$_{ISHE}$) measured in a pristine state of Bi$_{0.85}$La$_{0.15}$FeO$_3$ and corresponding PFM shown in \textbf{b}. $P$ with a yellow arrow in \textbf{a, e} indicates the direction of polarization fixed by the ferroelectric domain (strain state of the substrate) in two orientations of devices. No external electric field was applied during the experiment in \textbf{a, b, e, f}. The experimental protocol for pristine state magnon measurement is by changing the direction of the thermal gradient (or the magnon-propagation direction) by swapping the current/voltage electrode (schematic Figure \ref{fig:magnon1}, main text). $q$ represents the direction of the thermal gradient shown by the black arrow. Data in \textbf{a, e} and \textbf{c, g} is recorded under the same protocol, respectively. \textbf{c, g} is recorded under an external electric field where $q$ remains same only the direction of $P$ reverses. The corresponding PFM images after poling are shown in \textbf{d} and h for devices [100] and [010]. Single domain favors magnon propagation and hence the large inverse spin Hall voltage whereas multidomain does not allow magnon transmission. The power at the source electrode is fixed to 2mW corresponding to the $I_{ac}=$ 1.7mA.}
\end{figure*}

\begin{figure*}[htbp!]
\centering
\includegraphics[width=15cm]{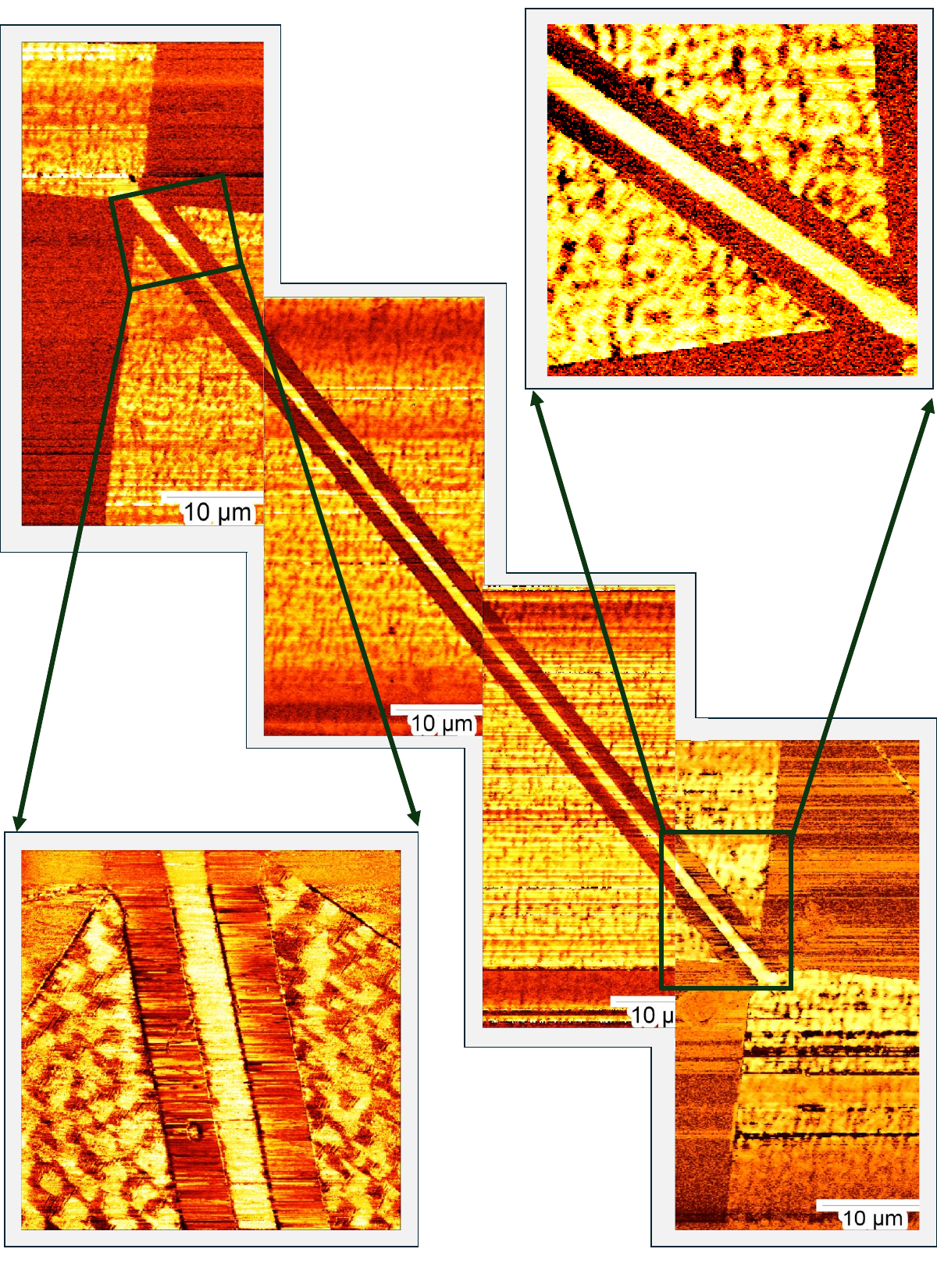}
\caption{\label{fig:Figure3} \textbf{Single ferroelectric domain:} Full device piezo force microscopy image to show a robust single domain formation after poling by the in-plane single electrical pulse (+150 kV/cm). The insets are the zoomed PFM scans to magnify the ferroelectric single domains in the two extreme edges of the 100$\mu$m long stripe of a nonlocal device. This PFM image belongs to one of $[010]$ devices as depicted in Fig.\ref{fig:magnon1}a.}
\end{figure*}

\begin{figure*}[htbp!]
\centering
\doublespacing
\includegraphics[width=18cm]{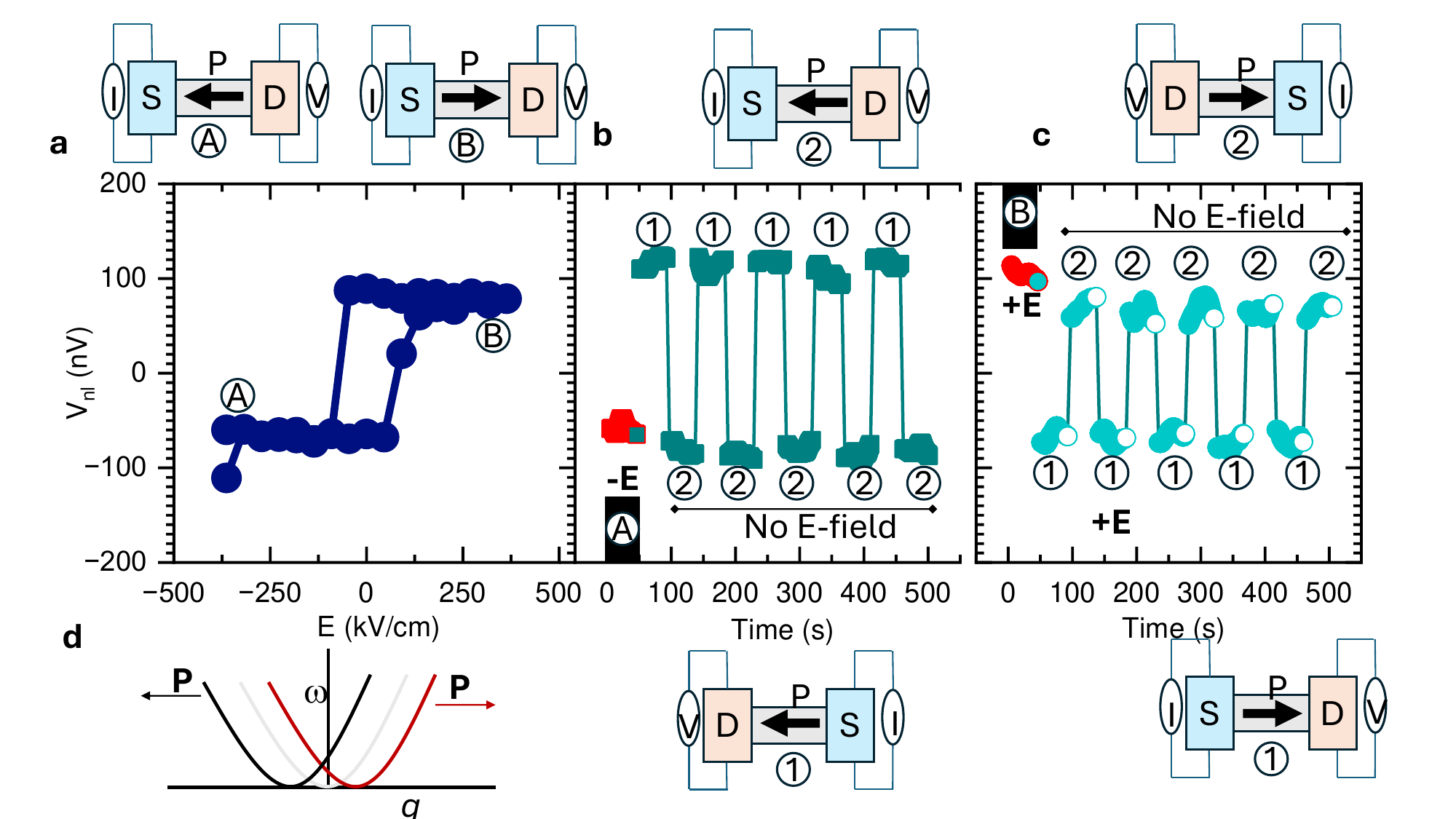}
\caption{\label{fig:RSM} \textbf{Polarization-driven non-reciprocal magnon transport:} \textbf{a} Magnon-electric field hysteresis, (1) and (2) represent the polarization state and source (S) and drain (D) correspond to the current source and voltage detector. \textbf{b}, the initial state is set by the electric field (in red), followed by the swapping of the electrode LEFT/RIGHT, changing the detector and source. This is done by automatically using the Keithley switch box, no physical movement of contact is involved therefore any non-reciprocity from the artifacts can be ignored. The circuit of states (1) and (2) are shown at the top and the bottom of the figure. Similarly, in the opposite polar situation in \textbf{c}, where the $P$ direction is set by the electric field and followed by the LEFT/RIGHT S and D swapping. The difference (reversed polarity and magnitude) between the two states (\textbf{b}, \textbf{c}) is due to the non-reciprocity in BLFO imposed by the polarization state as drawn in \textbf{d}. $P$, polarization, $\omega$ and $q$ represent the magnons' energy and propagation vector. Electrically, the change of the polar direction also reverses the DMI\cite{jorge_spin_cycloid_PRL,viretSkyrmionsinBFO}, which imposes the non-reciprocity in the system and hence the magnon signals are different in two opposite $P$ states, thus the polarization dependence is caused by the DMI.}
\end{figure*}

\begin{figure*}[htbp!]
\centering
\doublespacing
\includegraphics[width=16cm]{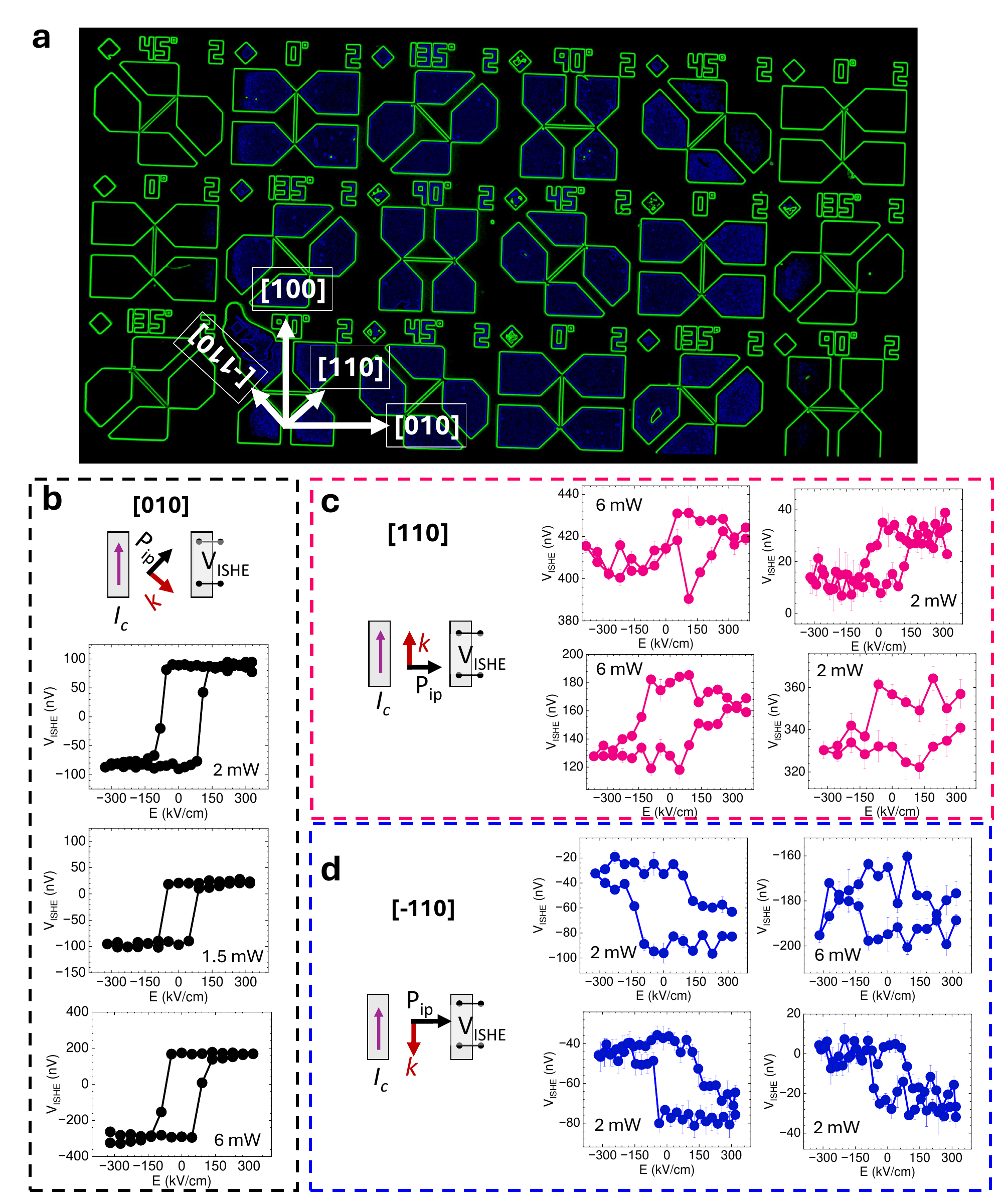}
\caption{\label{fig:fig5} \textbf{Magnon spin transport anisotropy:} \textbf{a} Optical image of patterned devices for spin transport measurements. 4-orientations were designed marked as 0$\degree$, 45$\degree$, 90$\degree$ and 135$\degree$ corresponding to pseudocubic directions [100], [110], [010], and [$\Bar{1}$10], respectively. Electro-magnon hysteresis recorded in devices \textbf{b}, [010], \textbf{c}, [110] and \textbf{d}, [$\Bar{1}$10]. $P_{ip}$ and $k$ represent the polarization and the propagation vector of the cycloid, respectively. Magnon hysteresis was recorded in different sets of devices on large-area samples. The schematics were drawn to represent the angle between the electrodes and the spin cycloid propagation vector. Irrespective of the magnitude of the nonlocal ISHE voltage (due to the different current injection) the hysteresis symmetry is preserved in three different device orientations. Device [010] is symmetric around zero whereas the ISHE voltage in [110] and [$\Bar{1}$10] all the way shows positive and negative minor loops, respectively. }
\end{figure*}

\end{document}


\title{\Large \centering Supplementary Information\\ 
Non-volatile spin transport in a single domain multiferroic}

\makeatletter
\renewcommand{\maketitle}{\bgroup\setlength{\parindent}{10pt}
\begin{flushleft}
  \textbf{\@title}

  \@author
\end{flushleft}\egroup
}
\makeatother

\onecolumngrid

\vspace{20cm}
\setstretch{-1.0}
\date{}
\author{%
\textsf{\textbf{Sajid Husain$^{1,	\dagger,*}$, Isaac Harris$^{1,2,\dagger,}$, Peter Meisenheimer$^{3}$, Sukriti Mantri$^{4}$, Xinyan Li$^{5}$ Maya Ramesh$^{6}$, Piush Behera$^{1,3}$, Hossein Taghinejad$^{2,7}$, Jaegyu Kim$^{3}$, Pravin Kavle$^{1,3}$, Shiyu Zhou$^{8}$, Tae Yeon Kim$^{3}$, Hongrui Zhang$^{1,3}$, Paul Stephenson$^{9}$, James G. Analytis$^{2}$, Darrell Schlom$^{6}$, Sayeef Salahuddin$^{3,10}$, Jorge Íñiguez-González$^{11,12}$, Bin Xu$^{13}$, Lane W. Martin$^{1,3,14,15}$, Lucas Caretta$^{8,16}$, Yimo Han$^{5}$, Laurent Bellaiche$^{4}$, Zhi Yao $^{1,*}$, Ramamoorthy Ramesh $^{1,2,3,14,*}$\\}
\textit{$^{1}$Materials Science Division, Lawrence Berkeley National Laboratory, Berkeley, CA, 94720, USA\\
$^{2}$Department of Physics, University of California, Berkeley, CA, 94720, USA\\
$^{3}$Department of Materials Science and Engineering, University of California, Berkeley, CA, 94720, USA\\
$^{4}$Physics Department and Institute for Nanoscience and Engineering, University of Arkansas, Fayetteville, Arkansas 72701, USA\\
$^{5}$Materials Science and NanoEngineering, Rice University, Houston, Texas, 77005, USA.\\
$^{6}$Department of Materials Science and Engineering, Cornell University, Ithaca, NY, 14850, USA\\
$^{7}$Heising-Simons Junior Fellow, Kavli Energy NanoScience Institute (ENSI), University of California, Berkeley, CA, 94720, USA\\
$^{8}$Department of Physics, Brown University, Providence, RI, 02906, USA\\
$^{9}$Department of Physics, Northeastern University, Boston, MA, 02115, USA\\
$^{10}$Department of Electrical Engineering and Computer Sciences, University of California, Berkeley, CA 94720, USA\\
$^{11}$Department of Materials Research and Technology, Luxembourg Institute of Science and Technology, 5 Avenue des Hauts-Fourneaux, L-4362 Esch/Alzette, Luxembourg\\
$^{12}$Department of Physics and Materials Science, University of Luxembourg, 41 Rue du Brill, L-4422 Belvaux, Luxembourg\\
$^{13}$Institute of Theoretical and Applied Physics, Jiangsu Key Laboratory of Thin Films, School of Physical Science and Technology, Soochow University, Suzhou 215006, China\\
$^{14}$Departments of Materials Science and 
NanoEngineering, Chemistry, and Physics 
and Astronomy, Rice University, Houston, TX, 77005, 
USA.\\
$^{15}$Rice Advanced Materials Institute, Rice 
University, Houston, TX, 77005, USA.\\
$^{16}$School of Engineering, Brown University, Providence, RI, 77005, USA\\}
{$^{*}$rramesh@berkeley.edu}\\
{$^{*}$jackie-zhiyao@lbl.gov}\\
{$^{*}$shusain@lbl.gov}\\
{$^\dagger$ These authors contributed equally}
}}
\date{\today}

\maketitle

\setcounter{figure}{0}
\renewcommand{\figurename}{\textbf{Supplementary Figure}}
\newpage
\tableofcontents

\newpage

\setstretch{2}
\begin{flushleft}\section*{SUPPLEMENTARY NOTE 1\\ X-RAY DIFFRACTION}
\end{flushleft}

The ferroelectric anisotropy in Bi$_{0.85}$L$_{0.15}$FeO$_3$ is expected due to the strain state of the orthorhombic substrate as a result of its monoclinic distortion. The representative $\theta-$2$\theta$ x-ray diffraction pattern of Bi$_{0.85}$L$_{0.15}$FeO$_3$(80nm) is shown in Supplementary Figure \ref{fig:FigureXRD}. The observation of only the $00l$ peaks validates the single phase epitaxial growth on DyScO$_3$ $(110)_O$. The misfit strain between Bi$_{0.85}$L$_{0.15}$FeO$_3$ and DyScO$_3$ (110)$_O$ is $\sim$0.2$\%$ and $\sim$0.33$\%$ along the $[001]_O$ and [1$\Bar{1}$0]$_O$, respectively. These values are evaluated using lattice constants measured using a reciprocal space map (RSM). DyScO$_3$ has an orthorhombic structure (space group $Pb\textit{nm}$, lattice constants, $a_0$=5.440 Å, $b_0$= 5.717 Å, and $c_0$=7.903 Å). For $[110]_O$ substrate, the orthorhombic unit cell can be related to the tilted pseudocubic (monoclinic) unit cell through the following steps, $a$=$c$$_0$/2=3.952 Å. The difference between the substrate lattice constant is $\sim$0.1$\%$ and between the angles is 3$\%$. In total, there is a large possibility of strain effect from the lattice difference as well as the monoclinic distortion of the substrate itself. We also identify the similar behavior in Pb$_{1-x}$Sr$_x$TiO$_3$ thin films deposited on DyScO$_3$ substrate \cite{kavle2023exchange}. A similar effect from the substrate is being used to create the anisotropic vortex tubes [PbTiO$_3$/SrTiO$_3$]$_n$ superlattices \cite{yadav2016observation}. Therefore, anisotropy (from the substrate) is expected in the ferroelectric domain wall motion vis-à-vis magnon transport under in-plane electric field excitation. The directions of Bi$_{0.85}$L$_{0.15}$FeO$_3$, DyScO$_3$, and how the metal electrodes landed on the top where the condition of switchable and non-switchable is depicted below.

\begin{figure*}[htbp!]
\centering
\includegraphics[width=16cm]{Figures/Supplementary Figure 1.pdf}
\caption{\label{fig:FigureXRD} \textbf{a} X-ray diffraction pattern of Bi$_{0.85}$L$_{0.15}$FeO$_3$.\textbf{b} Zoomed scan around BLFO $[001]$ peak. Oscillations in the Bi$_{0.85}$L$_{0.15}$FeO$_3$ peaks indicate good-quality epitaxy growth. \textbf{c} Atomic force microscopy image of Bi$_{0.85}$L$_{0.15}$FeO$_3$ surface where the atomic terrace (inset plot) indicates the layer-by-layer growth. \textbf{d} Schematic of Bi$_{0.85}$L$_{0.15}$FeO$_3$ unit cell (drawn using VESTA), DyScO$_3$ substrate along with the metal electrodes aligned at different angles $[100]$, $[010]$ and $[1\Bar{1}0]$ pseudocubic directions with respect to the substrate. The $[010]$ device electrodes are aligned with $[1\Bar{1}0]$ to the DyScO$_3$ orthorhombic notations. }
\end{figure*}
\clearpage
\newpage

\begin{flushleft}\section*{SUPPLEMENTARY NOTE 2\\ CROSS-SECTIONAL MICROSCOPY IMAGING AND POLARIZATION MAPPING}
\end{flushleft}

The polarization direction in BiFeO$_3$ is well understood which is along $<111>$ can choose either [111] [-111] [1-11] [11-1] [-1-11] [-11-1] [1-1-1] [-1-1-1]. From the atomic imaging, the projected atomic displacement (Methods, main text) is calculated to be 34.4$\pm$4.0 $pm$. The polarization vector direction is found to be 43.6$\degree$$\pm$3.8$\degree$. Thus in our case, the polarization of BFO (Supplementary Figure \ref{fig:FigureTEMBFO}) indeed is found set to be along [-1-11] or [1-11] as measured by the angle of polarization ($\sim$45$\degree$). The atomic displacement (real) using the projected displacement and considering the pseudocubic unit cell is calculated to be $\frac{34.4}{\sqrt{2}\times\sqrt{3}}$ = 14.04$\pm$1.5 $pm$.

\begin{figure*}[htbp!]
\centering
\includegraphics[width=16cm]{Figures/Supplementary Figure 20.pdf}
\caption{\label{fig:FigureTEMBFO} Cross-sectional microscopy and the polarization mapping of BiFeO$_3$. \textbf{a} Low magnification TEM image of the BiFeO$_3$/DSO. The high-resolution image \textbf{b} shows the polarization vector mapping. \textbf{c} Scheme to evaluate the polar distortion using atomic displacement. TEM gives the 2D projection however the magnitude of the displacement vector provides the direction of the polar distortion in the unit cell.}
\end{figure*}

In the La substituted BiFeO$_3$, the polarization is shifted away from $<111>$ and relaxed along $<112>$ \cite{yen-lin2020manipulating} and (Figure 1, theory, main text). From the STEM imaging (Supplementary Figure 3), the projected atomic displacement is found to be 32.4±3.5 pm, and using the pseudocubic unit the real atomic displacement is calculated to be $\frac{32.4}{\sqrt{5}\times\sqrt{6}}$ = 5.9$\pm$0.4 $pm$, which is smaller than the BiFeO$_3$ (discussed above) indicating the modifying polar distortion upon La-substitution. The polarization angle is evaluated to be 63.6±2.8$\degree $ $ \sim$63$\degree$. The 2D projection revealed the P-vector is downward choosing the [-1-1-2] or [1-1-2]. The magnitude and the direction confirm the polarization orientation is indeed along [112] in Bi$_{0.85}$L$_{0.15}$FeO$_3$.

\begin{figure*}[htbp!]
\centering
\includegraphics[width=16cm]{Figures/Supplementary Figure 19.pdf}
\caption{\label{fig:FigureTEMBLFO} Cross-sectional microscopy and the polarization mapping of Bi$_{0.85}$L$_{0.15}$FeO$_3$. \textbf{a} Low magnification TEM image of the Bi$_{0.85}$L$_{0.15}$FeO$_3$/DSO. High resolution \textbf{b} shows the polarization vector mapping. \textbf{c} Scheme to evaluate the polar distortion using atomic displacement. TEM gives the 2D projection however the magnitude of the displacement vector provides the direction of the polar distortion in the unit cell.}
\end{figure*}

\clearpage
\newpage
\begin{flushleft}
\textbf{SUPPLEMENTARY NOTE 3}\\
\textbf{Piezoresponse Force Microscopy \MakeUppercase{principle for ferroelectric domain imaging}}
\end{flushleft}

\begin{flushleft}\section*{SUPPLEMENTARY NOTE 3\\ \MakeUppercase{Piezoresponse Force Microscopy principle for ferroelectric domain imaging}}
\end{flushleft}

Piezoresponse Force Microscopy (PFM), is a scanning probe technique that leverages the inverse piezoelectric effect, elucidating the relationship between mechanical deformation in materials subjected to an applied electric field. In a typical Atomic Force Microscopy configuration, a minor AC voltage is applied between a conductive tip (here Pt) and a piezoelectric material (here Bi$_{0.85}$L$_{0.15}$FeO$_3$). This induces a local piezoelectric vibration, leading to the oscillatory deflection of the material surface. This deflection can be detected by the tip and transmitted to the lock-in system. The resulting amplitude signal provides information about the magnitude of the piezoelectric coefficient, while the phase signal identifies the polarization direction in the ferroelectric sample. PFM imaging unveils intrinsic ferroelectric domain structures by showcasing variations in spontaneous polarization across different domains through distinct PFM amplitude and phase signals.

To detect the polarization component oriented perpendicular to the sample surface, the vertical or out-of-plane (OP) PFM mode is employed. The tip, in contact with the sample surface, reflects the local piezoelectric response to the first harmonic component of tip deflection. The phase ($\phi$) provides insights into the local polarization directions beneath the tip. 

For detecting the polarization component within the plane of the sample surface, the lateral or in-plane (IP) PFM signal is recorded. The working principle of the lateral mode is grounded in a bias-induced surface shearing effect, with the detected polarization component sensitive to the direction perpendicular to the cantilever axis. The difference between the OP and IP modes is set up by choosing the desired oscillating frequency. For the OP case, the frequency of the cantilever is about 350kHz, and respectively for the in-plane is about 650kHz. Determining the two in-plane components of the polarization vector involves imaging the same region before and after a $\varphi=$90° rotation. The integration of an out-of-plane (OP) and two orthogonal in-plane (IP) PFM modes constitute the vector PFM approach. This combined methodology effectively reveals the three-dimensional contributions to the polarization vector in specific regions of a material.

Using the approach described above, one can easily distinguish the ferroelectric domain variant as depicted in Supplementary Figure \ref{fig:FigurePFMschematic}a. Individual domains have net polarization either IP or OP depending on the crystal structure. In the case of Bi$_{0.85}$L$_{0.15}$FeO$_3$ (used in this work), both IP and OP polarization are expected due to the rhombohedral (monoclinic) phase and polarization direction along $[111]$ or $[112]$ direction. Since the polarization is in diagonal to the unit cell, the in-plane domains are expected to be significant.
Let's go back to the domain variants in the Bi$_{0.85}$L$_{0.15}$FeO$_3$, which depend on the strain effect of the substrate. These variants are possibly in three types four, two, and single variants where each variant constitutes one domain. 
To probe the type of variant the sample can be rotated within the film plane (Supplementary Figure \ref{fig:FigurePFMschematic}a, bottom panel) and record the PFM signal. It shows how during the rotation, one can map out the polarization direction and hence number of variants.

Supplementary Figure \ref{fig:FigurePFMschematic} b and c, reveals the domain switched by $90\degree$ when the sample is rotated by $\phi=90\degree$. This is further substantiated in different devices as shown in Supplementary Figure \ref{fig:FigurePFM-multi-single}. This also shows the single variant in between the electrodes and two variants in the pristine state. How the two variants go into a single variant or single domain is demonstrated in Supplementary Figure \ref{fig:FigurePFM-hyst} \ref{fig:FigurePFM-multi-single} \ref{fig:FigurePFM-local} \ref{fig:FigurePFM_antiphase}. The polarization mapping is further discussed in the next section.

\begin{figure*}[htbp!]
\centering
\includegraphics[width=13cm]{Figures/Supplementary Figure 5.pdf}
\caption{\label{fig:FigurePFMschematic} \textbf{Ferroelectric domain variants and PFM imaging}. \textbf{a} Bi$_{0.85}$L$_{0.15}$FeO$_3$ unit cell and the four possible polarization variant $r_1$, $r_2$, $r_3$ and $r_4$. There are three possible situations in BFO compounds to be presented as four, two, and a single variant based on the ferroelastic domains. In four variants, four, 90$\degree$ domain walls are present, two variants most likely form 180$\degree$ domain wall whereas a single variant is free from the ferroelectric domain wall. The bottom panel in \textbf{a}, depicts the domain variants under in-plane rotation of the sample or the domain of different kinds. \textbf{b}  PFM images after poling and measured at $\varphi=0\degree$ and $\varphi=90\degree$ physically rotated sample. It appears that the polarization is rotated by $\varphi=90\degree$ indicative of the two variant systems under consideration.}
\end{figure*}

\begin{figure*}[htbp!]
\centering
\includegraphics[width=17cm]{Figures/Supplementary Figure 5_2.pdf}
\caption{\label{fig:FigurePFM-phi} \textbf{Lateral and Vertical PFM:} PFM phase maps of $[010]$ poled device recorded after physically rotated by 90$\degree$ and Lateral and vertical PFM images were recorded. A single-color contrast in all directions as well as in both PFM geometries indicates the single ferroelectric domain formation. The scale bar is 3$\mu$m.}
\end{figure*}

\begin{figure*}[htbp!]
\centering
\includegraphics[width=17cm]{Figures/Supplementary Figure 6.pdf}
\caption{\label{fig:FigurePFM-multi-single} \textbf{The evolution of single domain:} \textbf{a} PFM image of the pristine (unpoled) state of device $[010]$. \textbf{b} Ferroelectric hysteresis is recorded from 1-to-10 cycles and the evolution from multidomain to a single domain is represented by the squareness of the hysteresis evolution. \textbf{c} Corresponding PFM images of the single domain after poling.}
\end{figure*}

\begin{figure*}[htbp!]
\centering
\includegraphics[width=17cm]{Figures/Supplementary Figure 7.pdf}
\caption{\label{fig:FigurePFM-hyst} \textbf{The evolution of hysteresis in the device with multidomain and single domain devices:} Ferroelectric hysteresis in the device \textbf{a} $[100]$, \textbf{b} $[1\Bar{1}0]$ and \textbf{c} $[010]$. In the case of $[100]$, the ferroelectric state remains multidomain \textbf{(d)} whereas other devices form a single domain upon poling \textbf{(e) and (f)}. The ferroelectric polarization hysteresis has different features such as remnant polarization is lowest in $[100]$ and highest in $[1\Bar{1}0]$. This is correctly followed by the rule to $P_{[1\Bar{1}0]}=\sqrt{2}P_{[100] or [010]}$. Depending on the external electric field, the domain re-orientation is presented in \textbf{g $[100]$, h$ [1\Bar{1}0]$ and i $[010]$}. The domain wall motion leads to the formation of a single domain in two angles when the two domain polar vectors are parallel.}
\end{figure*}

\begin{figure*}[htbp!]
\centering
\includegraphics[width=14cm]{Figures/Supplementary Figure 8.pdf}
\caption{\label{fig:FigurePFM-local} \textbf{Piezoforce microscopy of $[100]$ device}: PFM images $[100]$ device after electrical ($E$) pulsing in two opposite directions. The big arrow indicates the field direction. Rectangles are used to identify the same area scanning in opposite poling. Rectangles within the metal stripes show the change of ferroelectric polarization locally as shown in zoomed schematics. Due to the non-regular domain pattern, the domains have some sections of head-to-head or tail-to-tail situations which create out-of-phase situations discussed in the next figure.}
\end{figure*}

\begin{figure*}[htbp!]
\centering
\includegraphics[width=16cm]{Figures/Supplementary Figure 9.pdf}
\caption{\label{fig:FigurePFM_antiphase} \textbf{Antiphase boundaries in multidomain regions}: \textbf{a} Lateral and \textbf{b} vertical PFM images of $[010]$ device after polling. The zoomed section of the vertical phase contrast represents the antiphase domain walls. \textbf{c,d} the antiphase DW and line scan shows the out-of-phase contrast around DWs.}
\end{figure*}

\clearpage
\newpage

\begin{flushleft}\section*{SUPPLEMENTARY NOTE 4 \\ SECOND HARMONIC GENERATION: POLARIZATION MAPPING} 
\end{flushleft}

Optical Second Harmonic Generation (SHG) involves the process of doubling the frequency of an incident light wave. The emission of SHG in a crystal is contingent upon its point group symmetry, making SHG responsive to any deviation from point symmetry in the material. As the development of ferroic order is closely associated with a reduction in crystal point-group symmetry, SHG serves as an effective means to investigate such ordered states in ferroelectric materials. Notably, this technique proves valuable for probing coexisting ferroic states or multidomain domains or different variants as we discussed above, in the multiferroics. The macroscopic description of the source term for SHG is described through the polarizability tensor \cite{BOYD20081,doi:10.1126/sciadv.abj5881},
\begin{equation}
    P_{i}^{2\omega} = \epsilon_{0} \chi_{ijk} E_{j}^{\omega} E_{kj}^{\omega}
\end{equation}
where $E^{\omega}$ is the incident electric field (laser light) at frequency $\omega$, $P_{i}^{2\omega}$ is the induced polarization in the nonlinear medium at frequency 2$\omega$, which acts as a source of an emitted, frequency-doubled, light wave with intensity $I_{SHG} \propto |Pi^{2\omega}|$. $\epsilon_{0}$ is the third-rank nonlinear susceptibility tensor, parameterizing the non-linear light-matter interaction. The indices $i, j, k$ are the components along $x, y, z$ as depicted in Supplementary Figure \ref{fig:FigureSHG1}a. The form of the $\chi_{ijk}$ tensor is dictated by the specific crystal point group symmetry. 
The nonlinear polarization at 2$\omega$ radiates back electrical field,
\begin{equation}
    \Delta^2E^{2\omega}-\epsilon_{0} \mu \frac{\partial^2 (E^{2\omega})}{\partial t^2}= \mu \frac{\partial^2 (P^{2\omega})}{\partial t^2}.
\end{equation}
The corresponding intensity will be,
\begin{equation}
    I^{2\omega} = \frac{1}{2} \epsilon_{0} \nu (E^{2\omega})^2.
\end{equation}
Experimentally, one can access the specific $\chi_{ijk}$ components by carefully selecting incident and detected light polarization or the angle of polarization by selecting $\varphi$. The spectral variation of the $\chi(2)$ tensor can be determined using a tunable wavelength light source, offering detailed insights into particular electronic transitions and optimizing second harmonic generation response. Figure \ref{fig:FigureSHG1}a illustrates a basic SHG setup being used in transmission mode in this work. The probe beam's polarization direction is defined by the polarizer angle, while the detected SHG light's polarization direction aligns with the analyzer angle. Since the second harmonic light is spectrally distinct from the fundamental light, monitoring the SHG frequency separately from the probe beam intensity simplifies the process, making it a background-free characterization technique.

In the context of ferroelectrics, the breaking of inversion symmetry is facilitated by a polar distortion, resulting in non-zero $\chi_{ijk}$ components. To comprehensively characterize this polar distortion along a specific crystallographic direction and distinguish between various tensor components, one can employ polarizer measurements. In these measurements, the polarization direction of the second harmonic light (analyzer angle) is fixed, while the polarization of the probe beam is systematically rotated.

During the experiment, it is crucial to account for three distinct coordinate systems and their interrelations: (i) the lab coordinate axes ($x, y, z$), which define the coordinate system within the optical setup; (ii) the sample coordinate axes ($X, Y, Z$), detailing the orientation of the sample edges in relation to the optics; and (iii) the crystal physics axes (X$_1$, X$_2$, X$_3$), representing the orientation of the nonlinear optical tensor coordinates for each domain variant (Supplementary Figure \ref{fig:FigureSHG1}b). Equation 3 encompasses polarization contribution directions aligned with the crystal physics axes of the 3$m$ point group symmetry.
\begin{figure*}[t]
\centering
\includegraphics[width=14cm]{Figures/Supplementary Figure 4_1.pdf}
\caption{\label{fig:FigureSHG1} \textbf{Second Harmonic Generation (SHG):} \textbf{a}  Typical SHG measurement setup in transmission. A laser (wavelength, 900nm) is used to excite the linear dichroic signal and harmonic signal detected by the photodiode detector. The incident probe beam is linearly polarized by a Glan-Taylor prism. The polarization of the probe beam is then set to an arbitrary polarization state by a rotatable half-wave plate. A half-wave plate for linear polarization states or a quarter-wave plate to achieve circular polarization. A focusing lens controls the spot size on the sample. The low-pass filter removes any second harmonic light generated in the polarization optics before the sample. The fundamental beam and higher-order harmonics are blocked by band/high-pass and low-pass filters. The SHG light is collected by the lens, and the polarization state is analyzed by projection on a second rotatable GT prism. The SHG intensity is integrated after passing a monochromator and photomultiplier tube. \textbf{b}  Schematics illustration of the axes orientation and polarization direction in the BLFO thin film sample. The transformation relations between the crystal physics axes (X$_1$, X$_2$, X$_3$) and the sample coordinate axes ($X, Y, Z$). The crystal physics axes (X$_1$, X$_2$, X$_3$) were determined from the 3$m$ point group symmetry. $\textbf{P}$ is in parallel to the [112]pc direction, as well as the $X_3$ of the crystal physics axes. In the SHG experiment, the fundamental light was incident normally onto the (001)pc surface as shown by red shaded line. }
\end{figure*}
The linear polarization of the incident beam is manipulated using a half-wave plate affixed to a motorized rotation stage. This allows precise control over the azimuth angle, $\varphi$, with respect to the x-axis in the $x-y$ plane of the ($x, y, z$) coordinate system. Consequently, the electric field of the fundamental light, expressed as $E^\omega (\varphi)$ = ($E_{X}, E_{Y}, E_{Z}$) = ($-E_{x}, -E_{y}, E_{z}$)= ($-E_{0} \cos{\varphi}, -E_{0} \sin{\varphi}, 0$), can be accurately rotated.
Here, $E_{1}, E_{2}, E_{3}$ are employed to denote the components of $E^\omega (\varphi)$ in the crystal physics axes ($X_1 \parallel [-110]$, $X_2 \parallel [-1-11]$, $X_3\parallel [112]$). The values of $\varphi$ determine specific light polarization states, allowing the determination of these components under varying conditions to enable the polarization measurement of different domains.

The light-induced non-linear polarization of the polarized domains can be described using the 3$m$ point group-based SHG tensor in a relation derived from Equation 1,

\begin{equation}
    \begin{pmatrix}
P_{1}\\
P_{2}\\
P_{3}\\
\end{pmatrix}
= 
\begin{pmatrix}
    0& 0&0&0&d_{31}&-d_{22}\\
    -d_{22}& d_{22}&0&d_{31}&0&0\\
    d_{31}& d_{31}&d_{31}&0&0&0\\
\end{pmatrix}
\begin{pmatrix}
    E_{12}\\
    E_{22}\\
    E_{32}\\
    E_{32}\\
    2E_{2}E_{3}\\
    2E_{1}E_{3}\\
    2E_{1}E_{2}\\
\end{pmatrix}
\end{equation}
where $d_{ij}$ ($i, j$ = 1,2,3) is the reduced nonlinear susceptibility tensor. The ($P_{1}, P_{2}, P_{3}$) corresponds 
to the polarization of SHG light (2$\omega$), whereas the ($E_{1}, E_{2}, E_{3}$) corresponds to the electrical field of fundamental light ($\omega$). By analyzing the fundamental light polarization settings, sample orientation, and nonlinear susceptibility tensors, we can derive valuable insights into intrinsic sample properties. These include the distribution of ferroelectric domains, as discussed in the main text and here, along with information about orientation and local structural symmetry.
The polarization components can be solved from the Equation 4 and written as,
\begin{align*} 
P_1^{2\omega}&= 2d_{13} E_1 E_3-2d_{22} E_1 E_2\\
P_2^{2\omega}&= -d_{22} E_1^2+d_{22} E_2^2+2d_{31} E_2 E_3 \\
P_3^{2\omega}&= d_{31} E_1^2+d_{31} E_2^1+d_{33} E_3^2\\
\end{align*}
The SHG intensity of polarization can be controlled by the incident light through the half-wave plate by selecting the angle $\varphi$. The SHG intensity achieved the maximum and the minimum at specific $\varphi$ with a 90° difference, for example, 45° and 135°, respectively. Due to the symmetry of the Bi$_{0.85}$La$_{0.15}$FeO$_3$, the in-plane polarization component $P_{net} \parallel$ diagonal of the substrate surface at either 45° or 135° with respect to $x$-axis, thus we can determine the SHG polarization map using the two extreme cases in parallel and perpendicular to in-plane $P_{net}$. We chose the two angles $\varphi=$45° and $\varphi=$135° polarization states of the fundamental light to measure the net in-plane polarization within the oppositely poled devices.
At $\varphi=$45° (light polarization $\parallel X_{1}$ or [$\Bar{1}$10]), the electric field component would be ($E_{1}, E_{2}, E_{3} = (E_{0}, 0, 0)$). And at $\varphi=$135° (light polarization $\perp X_{1}$ or [$\Bar{1}$10] and $\parallel$ [$\Bar{1}$ $\Bar{1}$ 0]), the electric field component would be ($E_{1}, E_{2}, E_{3} = (0, \frac{\sqrt{3}}{3} E_{0}, -\frac{\sqrt{6}}{3} E_{0})$).
The intensity in two orthogonal directions would be $I_x^{2\omega}\propto (\textbf{P}^{2\omega}.\textbf{e}_x)^2$, $I_y^{2\omega}\propto (\textbf{P}^{2\omega}.\textbf{e}_y)^2$. And total intensity is given by
\begin{align}
    I_{SHG}=I_x^{2\omega}+I_y^{2\omega}\propto [\,|P_1^{2\omega}|^2+|P_2^{2\omega}|^2+|P_3^{2\omega}|^2]\,.
\end{align}
We noticed that the intensity is maximum/minimum at $\varphi=$45° or $\varphi=$135° (Supplementary Figure \ref{fig:FigureS5}a and \ref{fig:FigureS5}b). We evaluate the difference between the two orientations to measure the real intensity from the polarization plotted in Supplementary Figure \ref{fig:FigureS5}c. This is also called a linear dichroism.  The polarization ($P_{net}$) intensity is maximum in the device [010] $\parallel$ $\varphi=$45° and hence $P_{net}$ is made and angle 45° to metal electrodes (Supplementary Figure \ref{fig:FigureS5}c, bottom) and no signal at $\varphi=$135°. This tends to explain the angle of polarization in the other devices after poling. For the domain variants, these results mean that SHG intensity would achieve higher values when the fundamental light polarization is parallel to the in-plane component of spontaneous polarization than when the fundamental light polarization is perpendicular to it. That is the reason why in these two cases the domain variants show a huge contrast in SHG mapping in the multidomain case (Supplementary Figure \ref{fig:FigureS5}c device [100], top panel).

\begin{figure*}[htbp!]
\centering
\includegraphics[width=16cm]{Figures/Supplementary Figure 4_2.pdf}
\caption{\label{fig:FigureS5} \textbf{Second Harmonic Generation (SHG):} Polarization maps recorded at \textbf{a} $\varphi$=45$\degree$ and \textbf{b} $\varphi$=135$\degree$ in three device orientations $[100]$, $[010]$ and $[1\Bar{1}0]$, respectively top, middle and bottom panel. \textbf{c} Differential map between \textbf{a} and \textbf{b} , i.e., $[I(45)-I(135)]/[I(45)+I(135)]$ .}
\end{figure*}

\begin{figure*}[t!]
\centering
\includegraphics[width=16cm]{Figures/Supplementary Figure 4.pdf}
\caption{\label{fig:FigureS6} \textbf{Second Harmonic Generation (SHG):} \textbf{a} \textbf{b} Polar map recorded on oppositely poled devices. Arrows represent the direction of the polarization in a single/multiple domain.}
\end{figure*}

\begin{figure*}[t!]
\centering
\includegraphics[width=16cm]{Figures/Supplementary Figure 4_3.pdf}
\caption{\label{fig:FigureS6} \textbf{Second Harmonic Generation (SHG):} \textbf{a} \textbf{b} Polar map recorded on oppositely poled devices. Arrows represent the direction of the polarization in a single/multiple domain.}
\end{figure*}
\clearpage

\newpage
\newpage
\clearpage

\begin{flushleft}\section*{SUPPLEMENTARY NOTE 5 \\ NITROGEN VACANCY MAGNETOMETRY}\end{flushleft}

Scanning nitrogen-vacancy (NV) magnetometry allows us to quantitatively detect the stray magnetic field from the very weak magnet such as compensated ferromagnet or antiferromagnetic thin layers \cite{finco2023single,rondin2014magnetometry} such as BiFeO$_3$ \cite{manuel2017real,meisenheimer2023persistent}. NV center defect comprises a nitrogen atom (N) substituting for a carbon atom and a vacancy (V) in one of the nearest neighboring sites within the diamond crystal lattice (Supplementary Figure \ref{fig:FigureNV1}a). This state features a spin triplet ($^3$A) ground level that can be initialized as depicted in Figure \ref{fig:FigureNV1}b, coherently manipulated, and read out solely through optical means at room temperature.

An integral characteristic of the NV defect pertains to its fundamental attribute: the ground state exists as a spin triplet state labeled $^{3}A_{2}$. This state's sub-levels undergo energy division due to spin–spin interaction, resulting in a singlet state with spin projection $m_{s} = 0$ and a doublet with $m_{s} = \pm1$. In the absence of a magnetic field ($B=0$), these $m_s=\pm1$ states are degenerate. The spin projection $m_{s}$ denotes the alignment along the intrinsic quantization axis of the NV defect, aligned with the crystal axis ($[111]$) joining nitrogen and the vacancy. Microwave excitation can be used to couple the $m_s=0$ and $m_s=\pm1$ states, allowing coherent manipulation of the spins. Optical excitation of the defect occurs through spin-conserving transitions to a spin triplet $^{3}E$ excited level. The $^{3}E$ level shares the same quantization axis and gyromagnetic ratio as the ground level. Upon excitation to the $^{3}E$ level, the NV defect can relax through either a radiative transition, resulting in broadband red photoluminescence (PL), or a non-radiative inter-system crossing to singlet states ($^{1}A$ and $^{1}E$).
\begin{figure*}[t]
\centering
\includegraphics[width=16cm]{Figures/Supplementary Figure 16.pdf}
\caption{\label{fig:FigureNV1} Nitrogen–vacancy (NV) centers in diamond. (a) An NV center is formed by a substitutional nitrogen atom and an adjacent vacancy in the diamond lattice. (b) The energy levels and optical transitions of the NV electron spin. The spin states can be polarized with a green laser and read out by fluorescence intensity. (c) Ground states of an NV center under an external magnetic field. The degeneration of states is lifted by the Zeeman effect. The hyperfine interaction with surrounding $^{13}$C nuclear spins brings an inhomogeneous broadening. The magnetic field is measured at around $\theta=53\degree$ from the NV axis. (d) Typical optical detected magnetic resonance spectrum of an NV center. The strength of the external magnetic field can be extracted from the resonant positions of the spectrum.}
\end{figure*}
These singlet states significantly influence the spin dynamics of the NV defect. Optical transitions, predominantly spin conserving ($m_{s} = 1$), coexist with non-radiative spin selective inter-system crossing to the $^{1}E$ singlet state, which exhibits strong spin selectivity. The shelving rate from the $m_{s} = 0$ sub-level is notably smaller than that from $m_{s} = \pm1$. Conversely, the NV defect tends to decay preferentially from the lowest $^{1}A_1$ singlet state to the ground state $m_{s} = 0$ sub-level. These spin-selective processes result in a non-thermal electron spin polarization into $m_{s} = 0$ through optical pumping. The photoluminescence intensity of the NV defect is significantly higher for $m_{s} = 0$ state is populated. Such a spin-dependent PL response enables the detection of ESR (electron spin resonance) on a single defect by optical means. Indeed, when a single NV defect, initially prepared in the $m_{s} = 0$ state through optical pumping, is driven to the $m_{s} = \pm1$ spin state by applying a resonant microwave (MW) field, a drop in the PL signal is observed. For magnetometry, the principle of the measurement is similar to the one used in optical magnetometers based on the precession of spin-polarized atomic gases. The applied magnetic field is evaluated through the detection of Zeeman shifts of the NV defect electron spin sub-levels. Indeed, when a magnetic field is applied in the vicinity of the NV defect (Supplementary Figure \ref{fig:FigureNV1}c), the degeneracy of $m_{s} = \pm1$ spin sub-levels is lifted by the Zeeman effect, leading to the appearance of two resonance lines in the ESR spectrum (Supplementary Figure \ref{fig:FigureNV1}d). A single NV defect therefore behaves as a magnetic field sensor with an atomic-sized detection volume.

The magnetic field is imprinted into the spectral position as $\Delta \nu$ of NV defects ESR. The relation between the ESR frequencies and the magnetic field can be understood from the ground state spin Hamiltonian of the NV defect, which is written as,
\begin{align}
    H= h [D S^2_z+E(S^2_x-S^2_y)]+g\mu_B \mathbf{B.S}
\end{align}
where $z$ is the NV defect quantization axis, $h$ is the Planck constant, $D$ and $E$ are the zero field splitting parameters, $S_x$, $S_y$ and $S_z$ the Pauli matrices, $g$=2.0 the Lande's $g$-factor, and $\mu_B$ the Bohr magneton. The $\mathbf{B}$ is the local magnetic field. The field along the NV axis can be considered as $|\mathbf{B_{NV}}|=|\mathbf{B}.\mathbf{u_{NV}}|$ with $\mathbf{u_{NV}}$ is NV center quantization axis. Therefore for any  $\mathbf{B_{NV}}$, the ESR frequency is described as,
\begin{align}
    \Delta \nu= 2 \gamma B_{NV}.
\end{align}
Where, $\gamma$=$g\mu_B/\hbar$ (28$\times10^9$ s$^{-1}$T$^{-1}$ with $g$=2). To determine magnetic field at a point the Zeeman splitting in the optically detected ESR spectrum is measured. The optimal response of the spin-dependent PL signal to a DC magnetic field is obtained by fixing a driving MW frequency to the maximal slope of a given ESR dip. Due to the change in the proximity magnetic field from the sample, the NV fluorescence rate $\frac{\partial I_0}{\partial B} \times \delta B \times \Delta t$ being $I_0$ is the PL intensity change and $\Delta t$ is the measurement time along with the photon-noise $\sqrt{I_0 \Delta t}$. Thus the field sensitivity is given as,
\begin{align}
    \eta &= \delta B \sqrt{\Delta t}\\ \nonumber
    &=\frac{\sqrt{I_0}}{\partial I/\partial B}\\ \nonumber
    &=\frac{4}{3\sqrt{3}} \frac{h}{g\mu_B} \frac{\Delta \nu}{C \sqrt{R_0}}.
\end{align}
Where $\Delta \nu$ is the ESR linewidth and $C$ is the ESR contrast. From the experimental limit presented by the Qnami \cite{Qnami}, the field sensitivity $\eta$ is expected to be 2.3$\pm$0.2$\mu$T/$\sqrt{T}$.
To record the measurements, the first step is to set up the tip-to-sample distance using the frequency modulation AFM mode where the $\Delta f$=15Hz ensures the distance $<5nm$ to keep the best sensitivity.  The microwave (MW) near-field antenna is brought in proximity ($<$50 $\mu m$) to the Quantilever.

There are two measurement modes possible in NV microscopy such as dual iso-B mode (where magnetic images exhibit two iso-magnetic-field (iso-B) contours \cite{tetienne2014nitrogen}), which allows for rapid visualization of the magnetic textures, and then the full-B mode, which allows for a full quantitative analysis of the stray field. 
The iso-B image is recorded at fixed MW frequency where the NV center experiences the magnetic field $B_{NV}(x,y)=B_0+B_{sample} (x,y)$ at resonance condition ($B_{NV}=\nu_{iso}/2\gamma$ and the MW induces information to NV spin and impact the fluorescence without information of the magnetic field from the sample.
The magnetic field information is further recorded at each pixel and the exact value is extracted from the splitting $\delta \nu$, which is known as Full-B.
The iso-B and full-B images are presented in several devices in Supplementary Figures \ref{fig:FigureSNV2}, \ref{fig:FigureNV3}, \ref{fig:FigureNV4}.

\begin{figure*}[htbp!]
\centering
\includegraphics[width=16cm]{Figures/Supplementary Figure 14.pdf}
\caption{\label{fig:FigureSNV2} \textbf{Iso-B and Full-B NV images}: \textbf{a} Topography \textbf{b} Iso-B o the film and  \textbf{c} full-B around the Pt electrode to measure the stray field. \textbf{d} Topography \textbf{e} Iso-B of the poled device and \textbf{f} full-B on the poled device. The data was recorded after poling in the device $[010] $(Figure 3a, top panel, main text). The magnetic texture is not clear behind the electrode due to screening from the Pt metal. However, the electric field line impact underneath the electrode is expected to be the same. Dotted lines indicate the edge of the Pt metal electrode on BLFO for electric field pulses. }
\end{figure*}
\newpage

\clearpage

\begin{figure*}[htbp!]
\centering
\includegraphics[width=16cm]{Figures/Supplementary Figure 15.pdf}
\caption{\label{fig:FigureNV3} \textbf{NV microscopy on different samples device for reproducibility check of the formation of the single variant in a single domain of Bi$_{0.85}$La$_{0.15}$FeO$_3$ device}: \textbf{a,b} Topography of the device $[1-10]$ and corresponding Iso-B NV image. \textbf{c,d} Topography of the device $[010]$ and corresponding \textbf{f} Iso-B NV image of the magnetic texture. A single variant due to the single domain formation is consistent in all kinds of devices of Bi$_{0.85}$La$_{0.15}$FeO$_3$ thin films.}
\end{figure*}
\newpage

\begin{flushleft}\section*{SUPPLEMENTARY NOTE 6 \\ SPIN CYCLOID WAVEVECTOR AND POLARIZATION RELATION IN POLED L\MakeLowercase{a}-BiFeO$_3$} \end{flushleft}

From the NV measurement, we discovered that the presence of random ferroelectric domains in Bi$_{0.85}$La$_{0.15}$FeO$_3$ results in a combination of magnetic phases, including cycloids and the G-type antiferromagnetic phase (refer to Figure 1 in the main text and Supplementary Figure 18, 19). The former illustrates distinct variants, indicating that the magnetic states are determined by the random ferroelectric domains. The in-plane net polarization, as observed through PFM, SHG (Figure 3), is perpendicular to the cycloid propagation vector, consistent with the BiFeO$_3$ parent compound. Consequently, the cycloid is anticipated to exhibit similar characteristics.

The choice of the cycloid's propagation vector—either a single variant or two variants (type-I or type-II cycloid)—depends on the type of ferroelectric domain wall in BiFeO$_3$, as well as the boundary conditions or strain effects, as discussed in ref. \cite{manuel2020antiferromagnetic}. On DyScO$_3$, BiFeO$_3$ prefers a bulk-like type-I cycloid with a two-variant orthogonal wave vector in 71$\degree$ ferroelastic domain walls. In the case of Bi$_{0.85}$La$_{0.15}$FeO$_3$ (this work), in poled devices between the electrodes (main text Figure 3), the spin cycloid follows the single domain and represents a single variant cycloid. Supplementary Figure \ref{fig:FigureSNV2} \textbf{a},\textbf{b} demonstrate the two possible polar states in the \textbf{BLFO} unit cell along with the possible spin cycloid propagation vectors. On projecting the combination of all the vectors, the $P$-$k$ orthogonal relation is only followed by the $[1-10]$ and $[110]$ based on the NV data Supplementary Figure \ref{fig:FigureSNV2} \textbf{c},\textbf{d}. Since the $P$ and $k$ are orthogonal thus the $[110]$ and [1-10] are only allowed vectors in Bi$_{0.85}$La$_{0.15}$FeO$_3$ poled devices.

\begin{figure*}[htbp!]
\centering
\includegraphics[width=16cm]{Figures/Supplementary Figure 17.pdf}
\caption{\label{fig:FigureNV4} \textbf{Spin cycloid propagation direction in Bi$_{0.85}$La$_{0.15}$FeO$_3$}: \textbf{a-b} Cycloid in two poled situations. The three possible spin cycloid propagation directions $[01\Bar{1}]$ $[\Bar{1}01]$ and $[1\Bar{1}0]$ when the polarization is along $[112]$. The lower panel indicates the $[1\Bar{1}0]$ is the suitable direction with relation to the polarization. When we mapped the same conditions on the NV images \textbf{c-d} the relation is testified that the $P$ and $k$ are orthogonal and $k$ is along $1\Bar{1}0$.}
\end{figure*}

\begin{figure*}[htbp!]
\centering
\includegraphics[width=16cm]{Figures/Supplementary Figure 2.pdf}
\caption{\label{fig:FigureS2} Polarization and cycloid direction switching under external electric field in the device $[100]$. The left panel depicts the two domains with two spin cycloid propagation vectors [$\Bar{1}$10] and$ [110]$ (main Fig. 2b, top panel). Right panel: after poling with electric field $E$, it rotates the $P_{ip}$ from $[112]$ to [$\Bar{1}$12] (71$\degree$ switch) and \textit{k} switches by 90$\degree$. This is due to the strain state of the DSO substrate and therefore we do not form a single domain in the $[110]$ direction.}
\end{figure*}

\begin{figure*}[htbp!]
\centering
\includegraphics[width=14cm]{Figures/Supplementary Figure 3.pdf}
\caption{\label{fig:FigureS3} Schematic indicates the direction of the polarization $P$, $P_{ip}$ under in-plane electric field $E$.}
\end{figure*}
\clearpage
\newpage

\begin{figure*}[htbp!]
\centering
\includegraphics[width=16cm]{Figures/Supplementary Figure 10.pdf}
\caption{\label{fig:FigureS12} \textbf{Polarization and wavevector relationship with the device orientation}: \textbf{a} $[010]$ and \textbf{b} $1\Bar{1}0$ and \textbf{c} $[110$]. In three cases, the Pt electrodes are rotated by $\pi/4$ with respect to the substrate edge $[001]_O$. We are not considering the case of $[100]$ due to multidomain formation and as a result, no magnon output is detected. In single-domain orientations, the single variant cycloids are formed. Here we show that the cycloid propagation vector $k$ has a fixed direction but due to changing the electrode orientation, the $k$-vector also moved with polar order as imaged in the main text Figure 3. }
\end{figure*}

\begin{figure*}[htbp!]
\centering
\includegraphics[width=17cm]{Figures/Supplementary Figure 11.pdf}
\caption{\label{fig:Figure13} \textbf{Mechanism of magnetic and polar order under electrical poling}: \textbf{a} Ferroelectric polarization with sublattice magnetization $m_1$, $m_2$. The magnetization lies in the $(112)$ plane where the $P$ is oriented along $[112]$ (Figure 1, main text). The in-plane projection of $P$ lies along $[\Bar{1}10]$, $[110]$ or [$\bar{1}$$\bar{1}$0] depending on the direction of the device and $P$ projection on the device. Schematic corresponding to the device orientation along \textbf{b} $[010]$ and \textbf{c} $[1\Bar{1}0]$ and \textbf{d} $[110]$. The cycloid chose the propagation direction orthogonal to $P$ \cite{manuel2017real,meisenheimer2023persistent} (Figure 3, main text), so in the case of $P_{110}$ cycloid chose the direction $k\parallel$[$\Bar{1}$10]. Similarly, the $P_{\Bar{1}10}$ allowed $k\parallel$$[1\Bar{1}0]$. In the polar region back and forth, the direction of $P$ is set by the electric field and depends on the device orientation. The $P$ is switched by $\pi$/2 in the device $[010]$ \textbf{(b)} where the direction of $k$ is also switched by $\pi$/2, this influences the sublattice magnetization and hence inverted upon poling, which shows the opposite ISHE voltage (Figure 4a $[010]$). For device $[1\Bar{1}0]$, $P$ is reversed oppositely which means the $k$ vector remain same [$\Bar{1}$10], however as in-plane $P$ switched by $\pi$ impact $m_1$, $m_2$ as depicted in \textbf{(c)}, which shows finite hysteresis. Similarly, the hysteresis device $[110]$ has the same but opposite effect to the device $[1\Bar{1}0]$.}
\end{figure*}

\clearpage

\begin{flushleft}\section*{SUPPLEMENTARY NOTE 7 \\ THERMALY EXCITED NON-LOCAL MAGNON TRANSPORT} \end{flushleft}

In our experimental geometry, we excite the magnon in the multiferroic using a thermal gradient generated by a Pt metal wire. Our primary focus lies on thermally excited magnons to gain insights into single-domain multiferroic physics and their influence on magnon transport. Specifically, we consider the second harmonic signal for analysis.
The Pt wire is heated by injecting the electrical current, and the resulting heat spreads radially (see Supplementary Figure 12a). Through the temperature gradient $\Delta T$ in the insulating magnet, the difference in magnon density between the hot and cold ends drives magnon diffusion from the hot end to the cold end. This phenomenon, governed by spin excitations or magnons, is known as the spin Seebeck effect (SSE). It's noteworthy that heat does not directly interact with magnons; rather, the temperature gradient generates phonons, and magnon-phonon scattering plays a significant role, especially at finite temperatures. All measurements are conducted at room temperature, ensuring homogeneous phonon contributions in various measurement categories. The magnon-phonon scattering contributes to relaxing magnon spin accumulation, linked to the Gilbert damping constant. Intriguingly, the thermally induced magnon excitations extend beyond the immediate source, spreading through the thermal gradient and possibly reaching several microns based on the temperature magnitude. 
In our study, we emphasize a relative comparison to comprehend magnon propagation in single-domain multiferroic materials. On the detector side, Pt is not considered a mere heat sink. As a consequence of the thermal gradient, a magnon chemical potential gradient forms (depletion at the injector and accumulation at the detector), initiating the transfer of magnon angular momentum beneath the detector.
Due to the spin-orbit coupling (SOC) of Pt (see Supplementary Figure 12b), an open circuit condition results in the generation of voltage due to the inverse spin Hall effect, the converse effect of the spin Hall effect is elucidated below. In the case of the second harmonic detection of the nonlocal voltage, the inverse spin Hall effect is used for electrical
detection of the spin Seebeck effect. Notably, no magnetic field is employed in this setup, ruling out any contributions from the Nernst effect or anomalous Nernst effect.

\begin{figure*}[htbp!]
\centering
\includegraphics[width=16cm]{Figures/Supplementary Figure 13.pdf}
\caption{\label{fig:FigureS14} \textbf{Thermal magnon transport}: \textbf{a} Nonlocal magnon transport geometry in multiferroics. \textbf{b} Spin Hall and inverse spin Hall effect. }
\end{figure*}

\newpage

\begin{figure*}[htbp!]
\centering
\includegraphics[width=16cm]{Figures/Supplementary Figure 21.pdf}
\caption{\label{fig:FigureS21} Nonlocal magnon transport in BiFeO$_3$ and BLFO recorded at identical conditions. BiFeO$_3$ data is shifted to (0,0) for better comparison. Bi$_{0.85}$La$_{0.15}$FeO$_3$ data is always symmetric around zero.}
\end{figure*}

\nocite{*}
\bibliography{apssamp_supp}